\documentclass[a4paper]{ar-1col}
\usepackage[numbers]{natbib}

\usepackage{amsmath,amssymb}
\usepackage{color}

\usepackage{mathrsfs}
\usepackage[x11names]{xcolor}
\usepackage[normalem]{ulem}

\jname{Annu. Rev. Nucl. Part. Sci.}
\jvol{AA}
\jyear{2019}
\doi{10.1146/annurev-nucl-101918-023510}

\begin{document}

\markboth{Murase and Bartos}{High-Energy Multi-Messenger Transient Astrophysics}

\title{High-Energy Multi-Messenger Transient Astrophysics}

\author{Kohta Murase$^{1,2,3,4}$ and Imre Bartos$^5$
\affil{$^1$Department of Physics, The Pennsylvania State University, University Park, PA 16802, USA; email: murase@psu.edu}
\affil{$^2$Department of Astronomy \& Astrophysics, The Pennsylvania State University, University Park, PA 16802, USA}
\affil{$^3$Center for Particle and Gravitational Astrophysics, Institute for Gravitation and the Cosmos, The Pennsylvania State University, University Park, PA 16802, USA}
\affil{$^4$Center for Gravitational Physics, Yukawa Institute for Theoretical Physics, Kyoto University, Kyoto, Kyoto 606-8502, Japan}
\affil{$^5$Department of Physics, University of Florida, Gainesville, FL 32611; email: imrebartos@ufl.edu}
}

\begin{abstract}
The recent discoveries of high-energy cosmic neutrinos and gravitational waves from astrophysical objects have led to the new era of multi-messenger astrophysics. 
In particular, electromagnetic follow-up observations triggered by these cosmic signals proved to be highly successful and brought about new opportunities in the time-domain astronomy.  
Here we review high-energy particle production in various classes of astrophysical transient phenomena related to black holes and neutron stars, and discuss how high-energy emission can be used to reveal the underlying physics of neutrino and gravitational-wave sources. 
\end{abstract}


\maketitle

\tableofcontents

\section{Introduction}
The new era of high-energy, multi-messenger astrophysics has begun by two recent breakthrough discoveries: (a)
the discovery of astrophysical high-energy neutrinos by the IceCube experiment in Antarctica~\cite{Aartsen:2013bka,Aartsen:2013jdh}; and (b) the direct detection of gravitational waves from the merger of two black holes by the Laser Interferometer Gravitational-Wave Observatory (LIGO)~\cite{Abbott:2016blz}. These detections have also been great triumphs of technological development for cosmic observations. 

The feasibility of time-domain multi-messenger astrophysics has been demonstrated by broadly coordinated observation campaigns in 2017 and 2018.   
This led to the discovery of gravitational waves from the neutron star merger event, GW170817, associated with the short gamma-ray burst (GRB), GRB 170817A, and the kilonova (also called Li-Paczynski nova or macronova) event, AT\,2017gfo~\cite{TheLIGOScientific:2017qsa,GBM:2017lvd}. The successful detection of electromagnetic counterparts at different wavelengths with follow-up observations strongly support the concordance picture of double neutron star mergers and short GRBs, and kilonova emission is consistent with heating by the decay of heavy radioactive nuclei.  
Another milestone detection was the high-energy neutrino event IceCube-170922A, with energy of $\sim0.1-1$~PeV. Follow-up observations revealed its association with a flaring blazar, TXS 0506+056, and enabled the determination of its multi-wavelength spectral energy distribution (SED), including the GeV-TeV gamma-ray band~\cite{Aartsen2018blazar1}. The interpretation of IceCube-170922A is still under debate, and confirmation by further observations will be important. 
Both of these success stories clearly demonstrate the potential in multi-messenger approaches, which combine information from different types of particles and waves (photons, neutrinos, gravitational waves, and cosmic rays), to reveal the origin of and the physical processes behind high-energy astrophysical phenomena.     

Neutrinos are elusive, neutral fermions. In the Standard Model there are three types of neutrinos, $\nu_e$ (electron neutrino), $\nu_\mu$ (muon neutrino), and $\nu_\tau$ (tau neutrino). They have tiny but finite masses, which was established by the observations of neutrino oscillation among the three generations. Besides gravity, they interact with matter only via the weak force. 
Consequently, gigantic detector volumes are required to detect astrophysical neutrino signals.  

In the MeV energy range, astrophysical neutrinos are mostly produced as a result of nuclear reactions, the best known examples being solar and supernova neutrinos. Neutrinos from the Sun were first measured by the HOMESTAKE experiment~\cite{1994PrPNP..32...13D}, which led to the rise of the solar neutrino problem|a large discrepancy between the then predicted and the measured neutrino flux from the Sun. The phenomenon of neutrino oscillation resolved this problem, and confirmed the standard model of the Sun. Supernova neutrinos were discovered in the wake of a nearby supernova, SN 1987A, by multiple water Cherenkov detectors around the globe (e.g., \cite{1992PhR...220..229K}). 

High-energy neutrinos, with energies beyond the GeV range, are produced by relativistic protons or ions, through hadronic interactions with matter or radiation. 
Detecting high-energy cosmic neutrinos is crucial to unravel the origin of cosmic rays --- which is one of the biggest mysteries in particle astrophysics. Cosmic rays are deflected by intergalactic magnetic fields, preventing us from pinpointing the location of their production site. High-energy photons can be produced by other mechanisms without involving cosmic-ray ions, and sufficiently high-energy gamma rays that are more likely to be hadronic are subject to electromagnetic interactions with lower-energy photons. These facts limit the use of the electromagnetic channel in probing cosmic ray sources. High-energy neutrinos serve as a more direct probe of cosmic particle accelerators, by which we can reveal their acceleration processes even in dense environments such as supernovae. 

Large-scale detectors are needed to detect ``high-energy cosmic neutrinos''. Currently operating and near future detectors include IceCube, a cubic-kilometer detector at the South Pole~\cite{Halzen:2016gng}; KM3Net, a cubic-kilometer detector under construction in the Mediterranean~\cite{Adrian-Martinez:2016fdl}, which is a successor of ANTARES in the Mediterranean~\cite{2011NIMPA.656...11A}; and the Baikal Deep Underwater Neutrino Telescope in Russia~\cite{1997APh.....7..263B}.

In contrast to high-energy particles that are accelerated in energetic outflows, gravitational waves are produced by the birth and dynamics of compact objects, in particular black holes and neutron stars~\cite{2011LRR....14....6S,2012LRR....15....8F,2013CQGra..30l3001B,2017RPPh...80i6901B}. 
Detectable gravitational waves require the non-axisymmetric acceleration of large amounts of matter, virtually ruling out any non-compact source. The strongest expected source of gravitational waves are the mergers of black holes and neutron stars. So far this is the only process from which gravitational waves have been detected~\cite{2018arXiv181112907T}. 
These mergers can emit a few percent of the rest mass of the merging objects in the form of gravitational waves, accounting for up to a few~$\times M_\odot$c$^2$ energy for stellar-mass black holes, and about $10^{-2}$~$M_\odot$c$^2$ for binary neutron stars. Gravitational-wave emission is weakly anisotropic, with the strongest emission along the binary's orbital axis being about 1.5 times higher than the emission in the average direction.
Another astrophysical process with sufficient matter and acceleration for substantial gravitational wave emission is stellar core collapse. During the collapse of massive stars, a neutron star can be formed. Gravitational waves are expected first from the violent collapse and the so-called bounce of the matter after it reaches neutron-star densities~\cite{2009CQGra..26f3001O, 2006RPPh...69..971K}
In the aftermath of the collapse, dynamical and dissipative instabilities can grow in the newly formed neutron star. 
In particular, if the progenitor star has high rotation, instabilities in the rapidly rotating neutron star can produce significant deviations from axisymmetry and produce copious amounts of gravitational waves~\cite{2013CQGra..30l3001B,2011GReGr..43..409A}. 
Depending on simulated models, gravitational waves from stellar core collapse is expected to be detectable for Galactic sources with advanced gravitational-wave detectors~\cite{2009CQGra..26f3001O}. 
However, if the rotational energy of the newly formed neutron star, or additional energy from fallback accretion, can be converted efficiently to gravitational waves, core collapse events could be detectable up to tens of megaparsecs with advanced detectors~\cite{2013CQGra..30l3001B}. 

The aim of this review is to summarize the current status and future prospects of multi-messenger particle astrophysics, focusing on transient sources of ``high-energy'' messengers. 
We begin with the introduction to high-energy particle production processes in the next Section. 
Then in Section \ref{sec:MMA} we give an overview of the current observational status and ongoing efforts of high-energy multi-messenger transient sources.  
In Section \ref{sec:models} we discuss different transient source models and our current understanding of the underlying emission processes in these events. We present a brief outlook for the future of the field and conclude in Section \ref{sec:outlook}.

\section{High-Energy Radiation Processes}
\subsection{Cosmic-Ray Acceleration}
Non-thermal emission is ubiquitous in astrophysical processes. 
The fact that cosmic rays are observed in a wide energy range from MeV to ultrahigh energies ($>$\,EeV) means that charged particles can gain energy by some process. 

Among various possible mechanisms, the most popular one is the Fermi acceleration mechanism, which was originally proposed by E. Fermi~\cite{Fermi:1949ee}. In this mechanism, charged particles gain energy stochastically via multiple interactions with scatters. 
Astrophysical shocks provide a viable setup for this type of particle acceleration to work. Ions can be reflected by magnetic fields at the shock. While the bulk of the particles are eventually advected to the far downstream, some of them can gain energy via scatterings by electromagnetic waves both in the upstream and downstream. In this diffusive shock acceleration mechanism~\cite{Drury:1983zz,1987PhR...154....1B}, a fraction of the energy of converging bulk flows can eventually be converted into the non-thermal energy of cosmic rays. 

The diffusive shock acceleration is not the only promising mechanism, and various mechanisms such as stochastic acceleration by turbulence and magnetic reconnections have been discussed in the literature. 
In any case, the particles have to be confined in the system, and the fundamental necessary condition is called the Hillas condition~\cite{Hillas:1985is}, which is,
\begin{equation}
\varepsilon < Ze r B (v/c), 
\end{equation}
where $\varepsilon$ is the particle energy, $Ze$ is the particle charge, $r$ is the system size, $B$ is the magnetic field strength, and $v$ is the characteristic velocity scale (that is the shock velocity for the diffusive shock acceleration mechanism). 
For relativistic sources, we have
\begin{equation}
\varepsilon< \Gamma Z e l' B',     
\end{equation}
where $l'$ is the comoving system size, $B'$ is the comoving magnetic field strength, and $\Gamma$ is the Lorentz factor. 
In reality, one has to take into account various cooling processes to evaluate the maximum energy of cosmic rays. But details depend on properties of the sources. 

If particle acceleration occurs in relativistic outflows such as GRB jets, the Hillas condition is rewritten as~\cite{Blandford:1999hi,Waxman:2008bj},
\begin{equation}
L_B > \frac{1}{2}\Gamma^2 c {\left(\frac{\varepsilon}{Ze}\right)}^{2}
\sim2\times{10}^{46}~{\rm erg}~{\rm s}^{-1}~\Gamma^2 {(\varepsilon/Z~{10}^{20.5}~{\rm eV})}^2 
\end{equation}
where $L_B$ is the magnetic luminosity of the outflow. 
This equation implies that accelerators of ultrahigh-energy cosmic rays (UHECRs) must be powerful. The number of candidate sources is rather limited, and the most promising ones are extragalactic transient sources such as GRBs and flares of active galactic nuclei (AGNs).

\subsection{Hadronic Processes}
High-energy cosmic rays interact with matter and radiation via hadronuclear and photohadronic interactions, respectively.  Hadronuclear interactions are mainly governed by inelastic $pp$ scatterings, in which neutrinos and hadronic gamma rays are produced via $p+p\rightarrow N(\pi^+,\pi^-,\pi^0)  + X \rightarrow N(\nu_\mu+\bar{\nu}_\mu+\nu_e+e^+, \nu_\mu+\bar{\nu}_\mu+\bar{\nu}_e+e^-, 2\gamma)+X$.
Interactions above baryon resonances are dominated by multi-pion production, leading to $\pi^+:\pi^-:\pi^0\approx1:1:1$, where $\pi^\pm$ is charged pion and $\pi^0$ is neutral pion. Although the inelastic $pp$ cross section gradually increases as energy, using the approximate constancy with $\sigma_{pp}\sim30$~mb and proton inelasticity with $\kappa_{pp}\sim0.5$, the effective optical depth to $pp$ interactions is given by   
\begin{equation}\label{eq:fpp}
f_{pp}[\epsilon_p]\approx\kappa_{pp}\sigma_{pp}c{t}_{\rm int}n_{N},
\end{equation}
where $t_{\rm int}$ is the interaction time and $n_N$ is the nucleon number density. For example, in the case of supernova shocks with size $r$ and velocity $v$, cosmic rays interact with target gas while they are confined and advected to the far downstream, so one expects $t_{\rm int}\approx r/v$. In the case of engine-driven supernovae, if cosmic rays from the engine travel through the ejecta almost rectilinearly, $t_{\rm int}\approx r/c$ is expected.   

Neutrinos and hadronic $\gamma$-rays can also be coproduced by the photomeson production, $p+\gamma\rightarrow N(\pi^+,\pi^-,\pi^0)  + X \rightarrow N(\nu_\mu+\bar{\nu}_\mu+\nu_e+e^+, \nu_\mu+\bar{\nu}_\mu+\bar{\nu}_e+e^-, 2\gamma)+X$,   
which is characterized by its effective optical depth, $f_{p\gamma}$. We consider a relativistic source with a target photon spectrum, $n_{\varepsilon'_t}$ (where $\varepsilon'_t\approx\varepsilon_t/\delta$ is the target photon energy in the comoving frame). Approximating the spectrum by ${\varepsilon'}_tn_{\varepsilon'_t}=n'_0{({\varepsilon'}_t/{\varepsilon'}_0)}^{1-\beta}$ with $\beta(>1)$ the power-law photon index and ${\varepsilon'}_0$ the reference energy, $f_{p\gamma}$ is given by~\citep[e.g.,][]{Murase:2015xka}
\begin{equation}\label{eq:fpgamma}
f_{p \gamma}[\varepsilon_p]\approx\eta_{p\gamma}[\beta]\hat{\sigma}_{p\gamma}l'n'_0
{(\varepsilon'_p/\tilde{\varepsilon}_{p\gamma0}^{\prime})}^{\beta-1},
\end{equation}
where $\eta_{p\gamma}[\beta]\approx2/(1+\beta)$, $\hat{\sigma}_{p\gamma}\approx\kappa_{p\gamma}\sigma_{p\gamma}\sim0.7\times{10}^{-28}~{\rm cm}^2$ is the attenuation cross section, $\bar{\varepsilon}_\Delta\sim0.3$~GeV, $\tilde{\varepsilon}_{p\gamma0}^\prime=0.5m_pc^2\bar{\varepsilon}_{\Delta}/\varepsilon'_0$, and $l'$ is the comoving size. 
This estimate is valid when the meson production is dominated by the $\Delta$-resonance and direct pion production. 

In either $pp$ or $p\gamma$ reaction, high-energy neutrinos are mostly produced as a result of pion and muon decay, and the neutrino energy fluence is written as,
\begin{equation}
E_\nu^2\phi_\nu \approx\frac{1}{4\pi d^2}\frac{3K}{4(1+K)}f_{pp/p\gamma} \frac{{\mathcal E}_{\rm cr}}{{\mathcal R}_{\rm cr}[\epsilon_p]},
\end{equation}
where $\phi_\nu$ is the neutrino fluence, $d$ is the distance to the source, ${\mathcal E}_{\rm cr}$ is the energy carried by cosmic rays, and ${\mathcal R}_{\rm cr}$ is a conversion factor from the bolometric energy to the differential energy of cosmic rays. Also, $K$ is a factor representing the ratio between charged pions and neutral pions, where $K\approx1$ and $K\approx2$ are for $p\gamma$ and $pp$ interactions, respectively.   
Realistically, pions and muons can be subject to various cooling processes, which modify resulting neutrino spectra. Thus, more generally, theoretical predictions for neutrino and gamma-ray spectra are model dependent. 

The meson production processes are among the ``hadronic processes'' that involve strong interactions. On the other hand, there are purely electromagnetic processes such as the Bethe-Heitler process and proton synchrotron radiation. The gamma rays originating from electromagnetic processes are also classified as hadronic components, because cosmic-ray ions are involved. 

\subsection{Leptonic Processes}
Gamma rays can be produced by leptonic processes as well as hadronic ones. Charged particles that relativistically move in magnetic fields emit synchrotron emission. The characteristic synchrotron energy is, 
\begin{equation}
\varepsilon_\gamma^{\rm syn}\approx 1.5\Gamma {\gamma'}_e^2 \hbar\frac{eB'}{m_ec}\sim200~{\rm eV}~\Gamma {(\gamma'_e/10^5)}^2(B'/1~\rm G),   
\end{equation}
where $\gamma'_e$ is the Lorentz factor of relativistic electrons in the comoving frame. For example, it is widely accepted that a low-energy component of the blazar SED is interpreted as synchrotron emission~\cite{2019Galax...7...20B}. 

High-energy electrons and photons interact via Compton scattering, $\gamma e\rightarrow\gamma e$. In the astrophysical context, gamma rays can be produced by the inverse-Compton process, in which low-energy photons gain energy via upscattering by relativistic electrons.
In the Thomson regime, where the photon energy in the electron rest frame is less than $m_ec^2$, we have
\begin{equation}
\varepsilon_\gamma^{\rm IC}\approx 2 {\gamma'}_e^2 \varepsilon_{\rm tar}\sim 20~{\rm GeV}~ {(\gamma'_e/10^5)}^2(\varepsilon_{\rm tar}/1~\rm eV),
\end{equation}
where $\varepsilon_{\rm tar}$ is the energy of target photons. The Klein-Nishina effect is important at sufficiently high energies. The cross section is suppressed when the photon energy in the electron rest frame exceeds $\sim m_ec^2$.
If target photons originate from synchrotron emission by primarily accelerated electrons, the process is called synchrotron self-Compton emission. 
If they originate elsewhere, the resulting emission is called external inverse Compton emission. In the case of blazars, the external photons can be provided by an accretion disk, broadline region, and dust torus~\cite{2019Galax...7...20B}. 

In the so-called leptonic scenario, observed gamma-ray emission is attributed to inverse-Compton radiation by primary electrons (or positrons). 
The electron luminosity and the magnetic field strength can be simultaneously determined through modeling of the SED.

\subsection{Electromagnetic Cascades}
Sufficiently high-energy gamma rays can interact with low-energy photons via the two-photon annihilation process, $\gamma\gamma\rightarrow e^+e^-$. 
Its optical depth is given by
\begin{equation}\label{eq:taugammagamma}
\tau_{\gamma\gamma}[\varepsilon_\gamma]\approx\eta_{\gamma\gamma}(\beta)\sigma_{T}l'(\varepsilon'_t n_{\varepsilon'_t})|_{\varepsilon'_t=m_e^2c^4/\varepsilon'_\gamma}\,,
\end{equation}
where $\sigma_T\simeq6.65\times{10}^{-25}~{\rm cm}^2$ is the Thomson cross section, $\varepsilon'_t$ is the target photon energy in the comoving frame, and $\eta_{\gamma\gamma}(\beta)\simeq7/[6\beta^{5/3}(1+\beta)]$ for $1<\beta<7$~\cite{Baring:2006bf}, which is the order of $0.1$. 
There is a correspondence between $p\gamma$ and $\gamma\gamma$ optical depths. The typical gamma-ray energy is given by $\varepsilon_\gamma\approx\Gamma^2m_e^2c^4{\varepsilon_t}^{-1}$ and we have~\citep[e.g.,][]{Murase:2015xka}
\begin{equation}\label{eq:opticaldepth}
\tau_{\gamma\gamma}[\varepsilon_\gamma^{c}]\approx\frac{\eta_{\gamma\gamma}\sigma_{\gamma\gamma}}{\eta_{p\gamma}\hat{\sigma}_{p\gamma}}f_{p\gamma}[\varepsilon_p]\sim{10}\left(\frac{f_{p\gamma}[\varepsilon_p]}{0.01}\right)\,,
\end{equation}
where $\varepsilon_\gamma^{c}$ is the gamma-ray energy corresponding to the resonance proton energy satisfying $\varepsilon_\gamma^c\approx2m_e^2c^2\varepsilon_p/(m_p\bar{\varepsilon}_\Delta)\sim{\rm GeV}~(\varepsilon_\nu/25~{\rm TeV})$.
The above equation implies that efficient emitters of $10-100$~TeV neutrinos are predicted to be ``dark'' as the sources of GeV-TeV gamma rays~\cite{Murase:2015xka}. Vice versa, GeV-TeV bright gamma-ray sources may not be ideal as the sources of neutrinos at $10-100$~TeV energies. 

If the intrasource $\gamma\gamma$ optical depth is larger than unity, high-energy gamma rays are attenuated inside the source. However, energy conservation implies that high-energy pairs produced via $\gamma\gamma\rightarrow e^+e^-$ keep generating lower-energy photons via synchrotron and inverse-Compton processes, which is called an electromagnetic cascade. The cascade can be induced by either primary ions or primary electrons, and it is important for powerful accelerators such as GRBs and blazars. Although details of an emergent spectrum depend on source parameters, a broad SED is formed as a generic trend, and the minimal proton-induced cascade fluence satisfies
\begin{equation}
\int^{\varepsilon_\gamma^{\rm S-cut}} d\varepsilon_\gamma \, (\varepsilon_\gamma \phi_\gamma^{\rm S-cas}) \approx \int_{0.5\varepsilon_\gamma^{\rm S-cut}} d\varepsilon_\nu \, \left(\frac{4+K}{3K} \varepsilon_\nu \phi_\nu\right),
\end{equation}
where $\varepsilon_\gamma^{\rm S-cut}$ is the energy at which the $\gamma\gamma$ optical depth is unity and $\phi_\gamma^{\rm S-cas}$ is the fluence of gamma rays cascaded inside the source. More generally, there are additional contributions to the cascade emission from the Bethe-Heitler and proton synchrotron processes. Efficient cascades are unavoidable in photon-rich sources, for which X-ray and gamma-ray observations are critical to examine bright neutrino sources. The relevance of intrasource cascades was demonstrated in the modeling of TXS 0506+056~\cite{Ahnen:2018mvi,Keivani:2018rnh,Murase:2018iyl,Cerruti:2018tmc,Gao:2018mnu}. 

Gamma rays capable of leaving the sources interact with cosmic radiation fields, including the cosmic microwave background and extragalactic background light. Except for ultrahigh energies, intergalactic cascades are governed by the two-photon annihilation and inverse-Compton scattering processes~\cite{Berezinsky:1975zz}. 
Note that if the target photon field is thermal the intrasource $\gamma\gamma$ optical depth decreases at high energies, allowing only high-energy gamma rays to escape from the sources~\cite{Murase:2015xka,Murase:2009ah}. 

It is known that the spectrum of intergalactic electromagnetic cascades is nearly universal, which is expressed as~\cite{Berezinsky:1975zz,Murase:2012df}
\begin{equation}
\varepsilon_\gamma^2\phi_\gamma^{\rm IG-cas}\propto
\begin{cases}
\varepsilon_\gamma^{1/2}
& (\varepsilon_\gamma\leq \varepsilon_\gamma^{\rm br})\\
\varepsilon_\gamma^{2-\beta}
& (\varepsilon_\gamma^{\rm br}<\varepsilon_\gamma<\varepsilon_\gamma^{\rm IG-cut})\\
\end{cases}
\end{equation}
where $\varepsilon_\gamma^{\rm IG-cut}$ is the cutoff energy due to the extragalactic background light, $\phi_\gamma^{\rm IG-cas}$ is the fluence of gamma rays cascaded in intergalactic space, $\varepsilon_\gamma^{\rm br}\approx 2{(\varepsilon_\gamma^{\rm IG-cut}/m_ec^2)}^2 \varepsilon_{\rm CMB}$, $\varepsilon_{\rm CMB}$ is the typical energy of the cosmic microwave background photons, and $\beta\sim2$ is the index that depends on details of cascades. For a TeV gamma-ray source located at a redshift of $z\sim1$, the cutoff due to the extragalactic background light typically lies in the 0.1~TeV range, which predicts a flat energy spectrum down to $\sim30$~MeV in the observer frame~\cite{Murase:2012df}. 

The multi-messenger connection among the diffuse cosmic particle (neutrinos, gamma rays, and cosmic rays) fluxes is crucial to reveal the origin of high-energy cosmic neutrinos. 
If IceCube neutrinos originate from inelastic $pp$ collisions and the sources are optically thin to $\gamma\gamma\rightarrow e^+e^-$ up to TeV energies (that are valid for star-forming galaxies and galaxy clusters), the fact that the isotropic gamma-ray background flux measured by Fermi-LAT~\cite{Ackermann:2014usa} is comparable to the diffuse neutrino flux leads to the conclusion that the intrinsic spectral index at the sources has to be $s<2.1-2.2$~\cite{Murase:2013rfa}. 

\section{Multi-Messenger Observational Status}
\label{sec:MMA}
\subsection{High-Energy Neutrino Observations and Electromagnetic Counterpart Searches}
The detection of high-energy cosmic neutrinos with PeV energies was first reported at Neutrino 2012 in Kyoto~\cite{Aartsen:2013bka}. The two events were found in the search for extremely high-energy neutrinos. The follow-up analysis on high-energy starting events led to the $4\sigma$ evidence of high-energy cosmic neutrinos~\cite{Aartsen:2013jdh}, and their existence has been established with accumulated data~\cite{Aartsen:2017mau}. Individual sources have not been firmly identified, so the IceCube flux can be regarded as the diffuse neutrino flux (or intensity). 
The measured diffuse neutrino flux is  $E_\nu^2\Phi_\nu\sim3\times{10}^{-8}~{\rm GeV}~{\rm cm}^{-2}~{\rm s}^{-1}~{\rm sr}^{-1}$ for all three flavors. These neutrinos consist of contributions from all sources that exist along the line of sight from the Earth, which is often called the astrophysical neutrino background. 
North-sky searches for track events induced by muon neutrinos have suggested a similar energy flux with a hard spectrum of $\Phi_\nu\propto E_\nu^{-2.1}$~\cite{Aartsen:2016xlq}. 
On the other hand, analyses on medium-energy starting events and shower events, which are sensitive to neutrinos below 100~TeV, have indicated a steeper spectrum, $\Phi_\nu\propto E_\nu^{-2.5}$~\cite{Aartsen:2017mau,Aartsen:2014muf}. 
The different spectral indices might indicate the existence of distinct components, and a large diffuse neutrino flux of  $E_\nu^2\Phi_\nu\sim{10}^{-7}~{\rm GeV}~{\rm cm}^{-2}~{\rm s}^{-1}~{\rm sr}^{-1}$ suggests a population of hidden neutrino sources owing to the tension with the isotropic gamma-ray background flux~\cite{Murase:2015xka}.  

Non-detection of point sources or high-energy ``multiplet'' sources (for which more than one neutrinos originate from a given position in the sky) implies that the source population responsible for the bulk of IceCube neutrinos is unlikely to be a rare class of astrophysical sources. Rather, abundant sources such as starburst galaxies, galaxy clusters/groups, and radio-quiet AGNs are favored. Next-generation detectors such as IceCube-Gen2 are essential to identify the main origin of IceCube neutrinos~\cite{Murase:2016gly}.
 
On the other hand, transient sources are detectable with the current IceCube if a bright burst or flare occurs. Time- and space-coincidence also allows us to significantly reduce atmospheric backgrounds. 
This advantage has been exploited for stacking searches for neutrino emission from GRBs. The non-detection of coincident events between neutrinos and GRBs have led to important constraints on cosmic-ray acceleration in GRBs~\cite{Abbasi:2012zw,Aartsen:2014aqy}. 
Stacking searches for neutrino-supernova associations have also been done~\cite{Senno:2017vtd,Esmaili:2018wnv}. 
Multiplet searches are also powerful for the transient neutrino sources~\cite{Aartsen:2017snx,Aartsen:2018fpd}. 

\begin{figure}[th]
\includegraphics[width=3in]{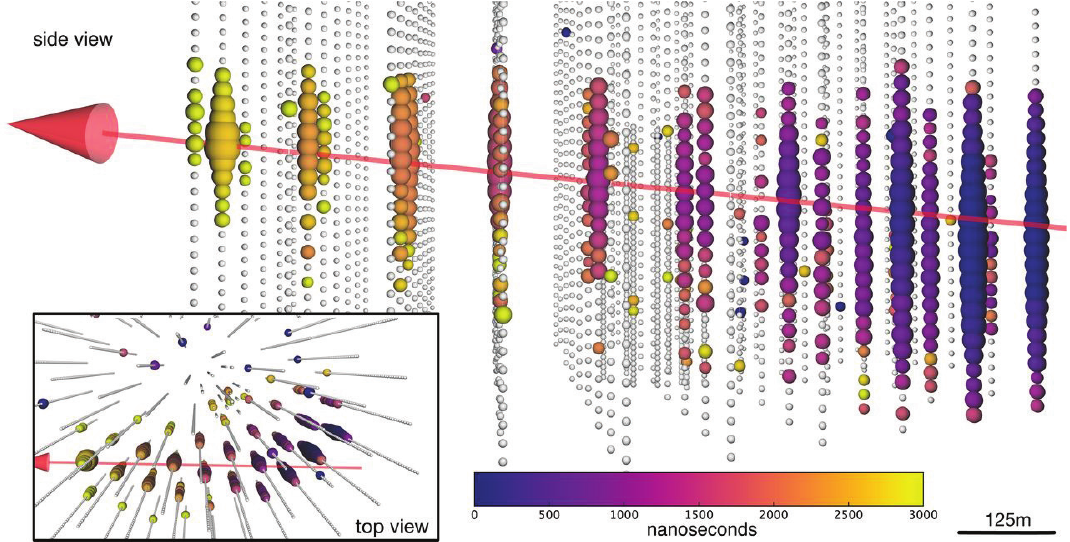}
\includegraphics[width=3in]{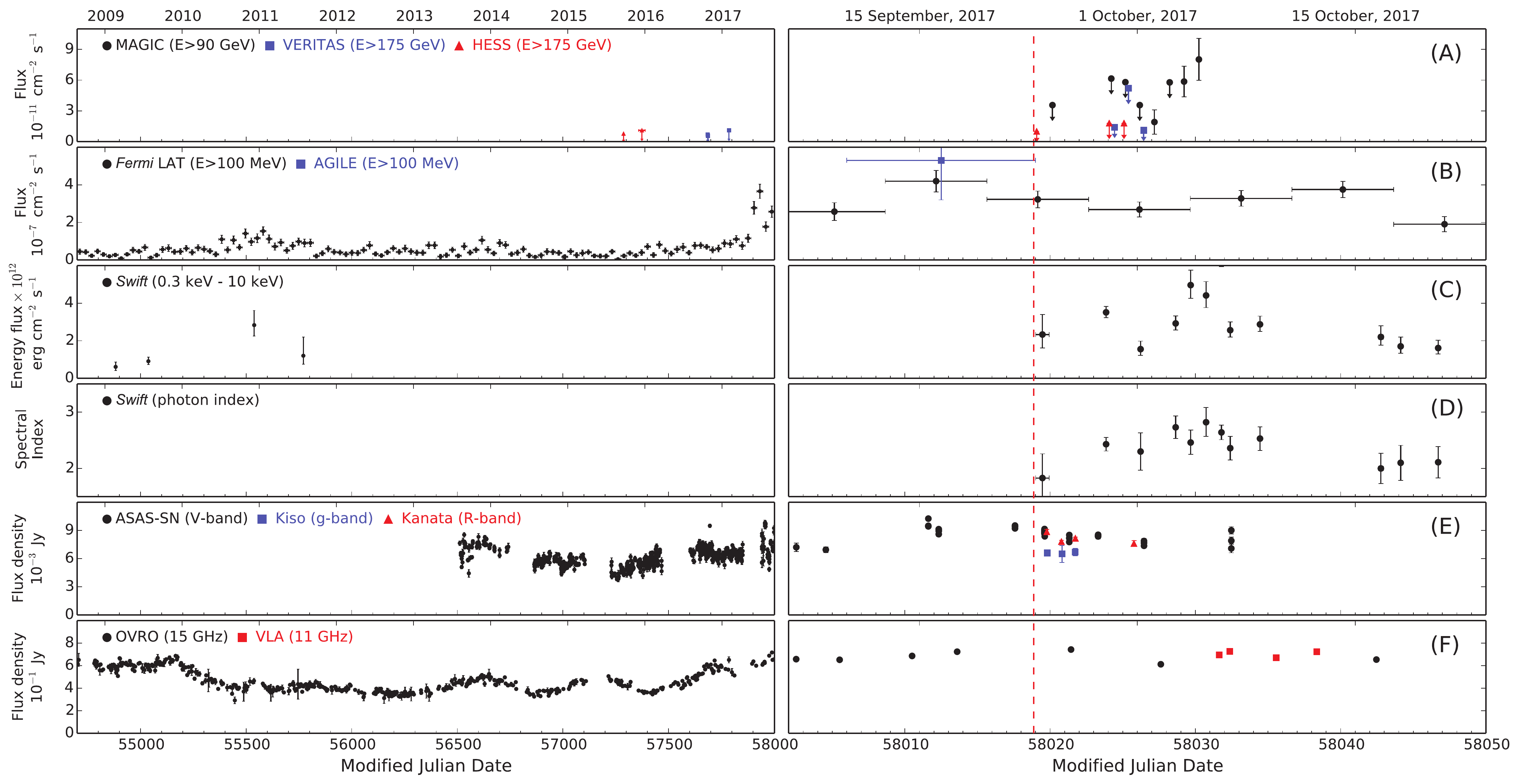}
\caption{Multi-messenger observations of TXS 0506+056 associated with the high-energy neutrino event IceCube-170922A. (a) Image of the neutrino-induced track event. (b) Multiwavelength light curves. The vertical dashed line indicates the timing of the detection of IceCube-170922A. Figure adapted from Ref.~\cite{Aartsen2018blazar1}.}
\label{fig1}
\end{figure}

Neutrino-triggered follow-up observations provide an alternative way of identifying the sources of high-energy neutrinos~\cite{Murase:2006mm,Kowalski:2007xb}. The real-time alert system in IceCube was developed for this purpose. To dig out subthreshold multi-messenger signals, Astrophysical Multimessenger Network Observatory (AMON) has attempted to combine multi-messenger information in a real-time manner~\cite{Smith:2012eu}. 
The feasibility of such a neutrino-triggered follow-up approach was best demonstrated in the observations of TXS 0506+056 that coincided with IceCube-170922A~\cite{Aartsen2018blazar1}, as presented in Fig.~\ref{fig1}. 
Within the error circle of this high-energy neutrino event, several blazar candidates in Kanata Telescope were identified, and one of them turned out to be a Fermi-LAT blazar in the high state. Several X-ray sources were identified by Swift follow-up observations, and TXS 0506+056 was further observed by NuSTAR. This source was also seen by the MAGIC gamma-ray telescope.  
The significance of the association with the gamma-ray flare was $\sim3\sigma$, which is insufficient to claim the discovery. However, interestingly, archival analyses on the past track data in TXS 0506-056 revealed the neutrino flare in 2014-2015~\cite{Aartsen2018blazar2}. Although associated gamma-ray flares were not found for this past neutrino flare event, the $\sim4\sigma$ significance gave us intriguing evidence for this blazar as a potential neutrino source.  

\subsection{Gravitational Wave Observations and Electromagnetic Counterpart Searches}
Gravitational-wave searches have been used to initiate and to follow-up electromagnetic observations since the era of first-generation gravitational wave detectors began more than a decade ago~\cite{2004CQGra..21S.765M,2005NuPhS.138..446M,2007AAS...211.9903P,2008ApJ...681.1419A,2012A&A...541A.155A,Abadie:2012bz}. With the limited sensitivity of initial gravitational-wave detectors, which were able to detect binary neutron star mergers out to about 20\,Mpc, a common detection was possible but not probable. Nevertheless, gravitational-wave searches triggered by electromagnetic observations resulted in a few astrophysically meaningful constraints, for example by indicating that a short gamma-ray burst GRB\,070201 directionally coincident with M31 could not have been produced by a compact binary merger~\cite{2008ApJ...681.1419A}.

Gravitational-wave observations with advanced detectors brought about a wide-scale, broadband electromagnetic follow-up effort. The first gravitational-wave discovery of binary black hole merger GW150914 on September 14, 2015, was followed up by over 60 observing facilities covering radio, optical, near-infrared, X-ray, and gamma-ray wavelengths~\cite{2016ApJ...826L..13A}. 

One of these facilities, the Gamma-ray Burst Monitor on the Fermi Satellite, reported the detection of a spatially and temporally coincident, albeit marginal, short gamma-ray burst 0.4\,s after the binary merger~\cite{2016ApJ...826L...6C,2016ApJ...827L..34V}, albeit the significance of this detection is debated~\cite{2016ApJ...827L..38G}. A marginal short-GRB counterpart was also detected later for another binary black hole merger discovered through gravitational waves, GW170104, in this case by the AGILE satellite~\cite{2017ApJ...847L..20V}. This GRB was, however, not observed by other detectors~\cite{2017ApJ...846L...5G}, and an unrelated, directionally overlapping long GRB complicated matters~\cite{0004-637X-845-2-152}. 

The electromagnetic follow-up campaign to identify gravitational waves finally triumphed with the observation of binary neutron star merger GW170817~\cite{TheLIGOScientific:2017qsa,GBM:2017lvd,2017ApJ...848L..13A}. This merger was discovered simultaneously by LIGO/Virgo and by Fermi-GBM, the former through gravitational waves within minutes and the latter through its short GRB counterpart within seconds after the event~\cite{2017ApJ...848L..13A,2017ApJ...848L..14G}. The spatial and temporal overlap between these two detections was rapidly recognized, and initiated a broadband, multi-messenger search for emission from the source. 

\begin{figure}[th]
\includegraphics[width=3in]{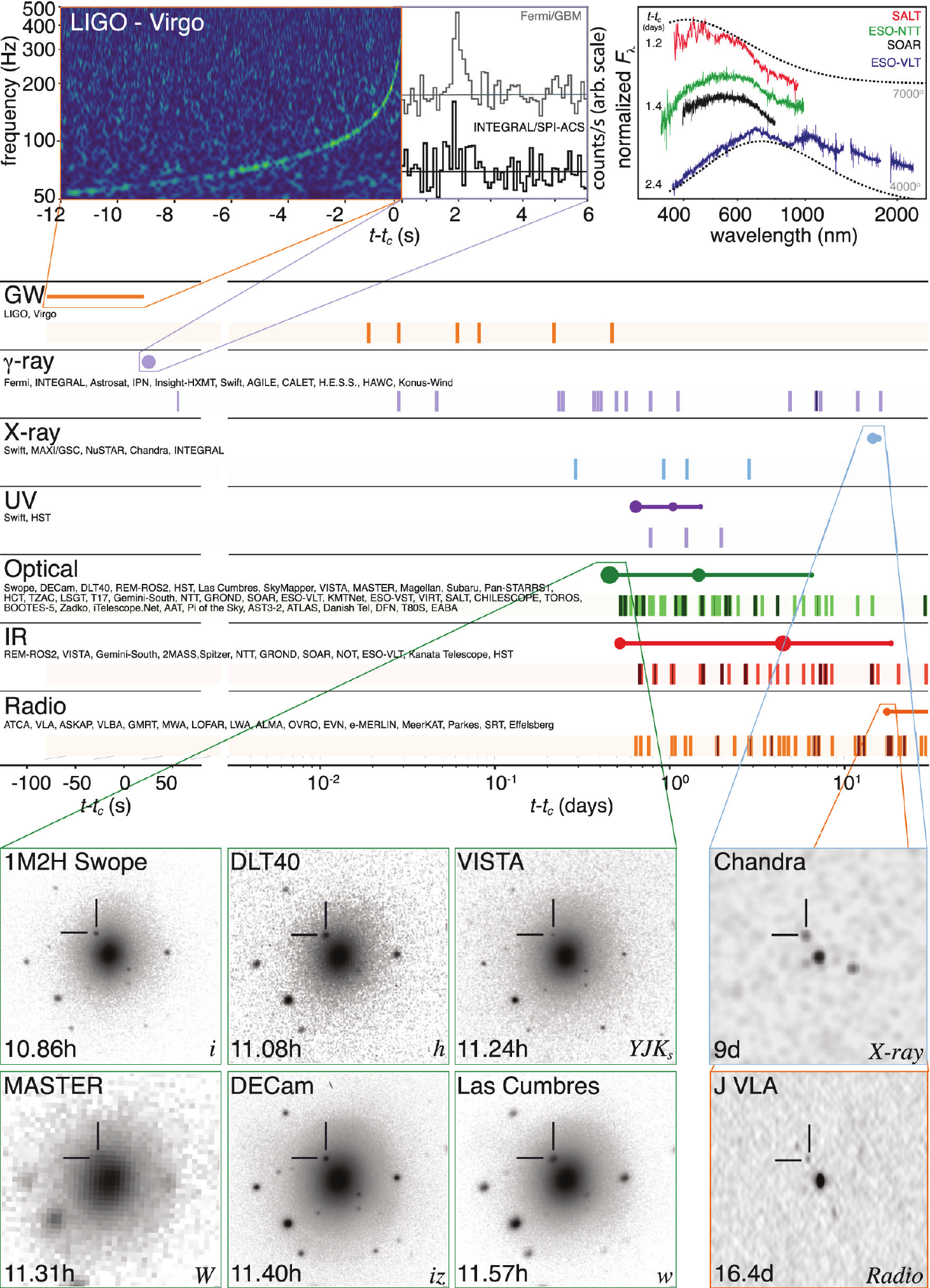}
\caption{Multi-messenger observations of GW170817. Figure adapted from Ref.~\cite{GBM:2017lvd}.}
\label{fig2}
\end{figure}

Fig.~\ref{fig2} presents a visual summary of the follow-up effort to detect GW170817 / GRB\,170817A. Less than 11 hours after the merger, its optical kilonova emission was found, first by the Swope Telescope~\cite{2017Sci...358.1556C}. X-ray and radio emission from the GRB afterglow was detected only with a large delay; 9 and 16 days after the merger, respectively, by the Chandra X-ray telescope and by the Jansky Very Large Array~\cite{XrayGW170817,2017Sci...358.1579H}.

GW170817/GRB\,170817A provided a wealth of unique information on high-energy emission from binary neutron star mergers that changed our GRB paradigm, and that would have not been possible without the observation of both gravitational waves and electromagnetic emission. First, the binary merger occurred at a large inclination angle of $15^\circ-40^\circ$~\cite{2018arXiv180511579T}. Prior to this discovery, GRB observations were only anticipated for smaller angles. 
Second, detailed afterglow observations further revealed that the outflow is structured, with a narrow energetic component along the orbital axis, and weaker emission at greater inclinations~\cite{2018MNRAS.479..588G,2018PTEP.2018d3E02I,2018Natur.554..207M,2018PhRvL.120x1103L,2018ApJ...856L..18M,2018MNRAS.tmpL..60T,2018Natur.561..355M}. 
Such a structured emission was not incorporated previously in GRB population studies prior to GW170817. This could mean that either GRB\,170817A is a rare event type~\cite{2019MNRAS.483..840B}, or a population of nearby GRBs observed at high inclination angles and without reconstructed distances may exist~\cite{2017arXiv171005931M,2018arXiv180806238G,2018arXiv181111260B}. This would also mean that a non-negligible fraction of future gravitational-wave observations of binary neutron star mergers will be accompanied by detectable GRBs, promising frequent high-energy multi-messenger discoveries.

\subsection{Coincidence Searches for Gravitational Waves and Neutrinos}
The search for common sources of gravitational wave and neutrinos has a long history, going back to the first extrasolar neutrino source, SN 1987A~\cite{1992PhR...220..229K,1989NCimC..12...75A}. In the era of interferometric gravitational-wave observatories, joint searches were investigated in detail starting around 2006~\cite{2008CQGra..25k4039A,2009NIMPA.602..268P,2009IJMPD..18.1655V,2009PhRvL.103c1102P,2011APh....35....1B,2012PhRvD..85j3004B,2018arXiv181011467B}. No such joint detection has been made so far, making this the next frontier in the multi-messenger puzzle.

This search type consists of two distinct categories, based on the type of the neutrino signal. In the first category are sources producing non-thermal, high-energy neutrinos with GeV energies and beyond, while in the second category are thermal, MeV neutrinos. In this review we will restrict our discussion to the first category, which is closely connected to high-energy emission.

The first searches for common sources of gravitational waves and high-energy neutrinos were carried out with the Initial LIGO and Virgo detectors, and with the partially completed IceCube and ANTARES detectors~\cite{2011PhRvL.107y1101B,2013JCAP...06..008A,2014PhRvD..90j2002A}. These analyses targeted event candidates for which neither the gravitational-wave nor the neutrino data was sufficiently significant to confidently indicate an astrophysical signal. 

Advanced LIGO's first observing run from September, 2015 until January, 2016 brought about the first gravitational-wave detections~\cite{2016PhRvX...6d1015A}, and with them the first targeted searches for high-energy neutrinos from established gravitational-wave sources using the IceCube, ANTARES and Pierre Auger observatories~\cite{2016PhRvD..93l2010A,2017PhRvD..96b2005A,2016PhRvD..94l2007A}. All three binary black hole mergers from this period, discovered via gravitational waves, were followed up by neutrino searches. Neutrino emission from these events was constrained to isotropic-equivalent energies less than $\sim10^{51}-10^{54}$\,erg, assuming neutrino spectrum $dN/dE\propto E^{-2}$. The spread in this emission constraint is due to the large localization uncertainty of gravitational waves, as the sensitivity of neutrino detectors can significantly change over the source directions allowed by gravitational waves. 

\begin{figure}[th]
\includegraphics[width=3in]{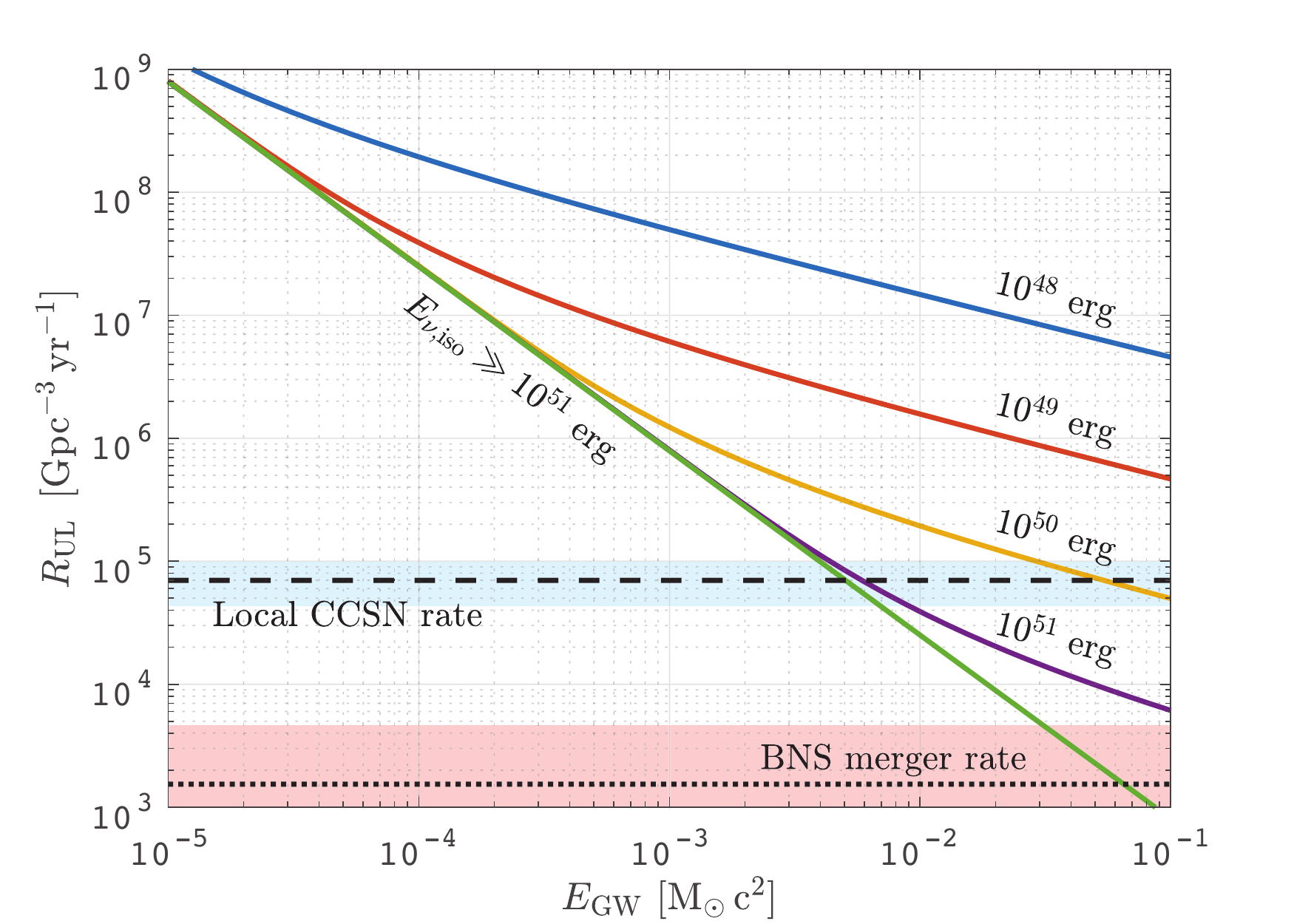}
\caption{Constraints on rate density of high-energy neutrino transients as a function of energy released as gravitational waves. Here $E_{\nu, \rm iso}$ is released energy of neutrinos and $E_{\rm GW}$ is released energy of gravitational waves. 
Abbreviations: BNS, binary neutron star; CCSN, core-collapse supernova.
Adapted from Ref.~\cite{2018arXiv181010693A}.}
\label{fig:O1neutrino}
\end{figure}

Beyond highly significant discoveries, gravitational-wave and high-energy neutrino data during Advanced LIGO's first observing run was also analyzed in search of events that remained below the detection threshold in individual data channels \cite{2018arXiv181010693A}. While no joint event was discovered, this search represented a sensitivity improvement of more than two orders of magnitude over previous similar searches carried out with earlier-generation detectors. Fig.~\ref{fig:O1neutrino} shows the observational constraints derived from this analysis. For realistic source rates of $<10^5$\,Gpc$^{-3}$yr$^{-1}$, these constraints limit the source population in the strong emission regime of gravitational-wave energy ${\mathcal E}_{\rm GW}\gtrsim10^{-2}$\,M$
_{\odot}$c$^{2}$ and isotropic-equivalent neutrino energy ${\mathcal E}_{\nu}\gtrsim10^{51}$\,erg. 

During Advanced LIGO/Virgo's second observing run from November, 2016 until August, 2017, coincident neutrinos were searched at near-real time with the IceCube and ANTARES neutrino observatories, and over a period of about a day with the Pierre Auger Observatory, following every gravitational-wave detection~\cite{2017ApJ...850L..35A,2017EPJC...77..911A}. This rapid analysis was motivated by the fact that a coincident neutrino would significantly aid to electromagnetic observations. While gravitational-wave localizations are typically limited to hundreds of square degrees~\cite{2016LRR....19....1A}, high-energy neutrinos can be reconstructed to sub-degree precision. This substantially reduces the number of electromagnetic foreground transients and the sky area that observatories need to survey to identify electromagnetic emission from the source. Since both gravitational waves and neutrinos are expected to be emitted by the main sources of interest within minutes~\cite{2011APh....35....1B}, multi-messenger identification on a similar time scale can aid the search for electromagnetic emission, such as a gamma-ray burst afterglow or kilonova, which can be observable over a longer period. 

\begin{figure}[th]
\includegraphics[width=3in]{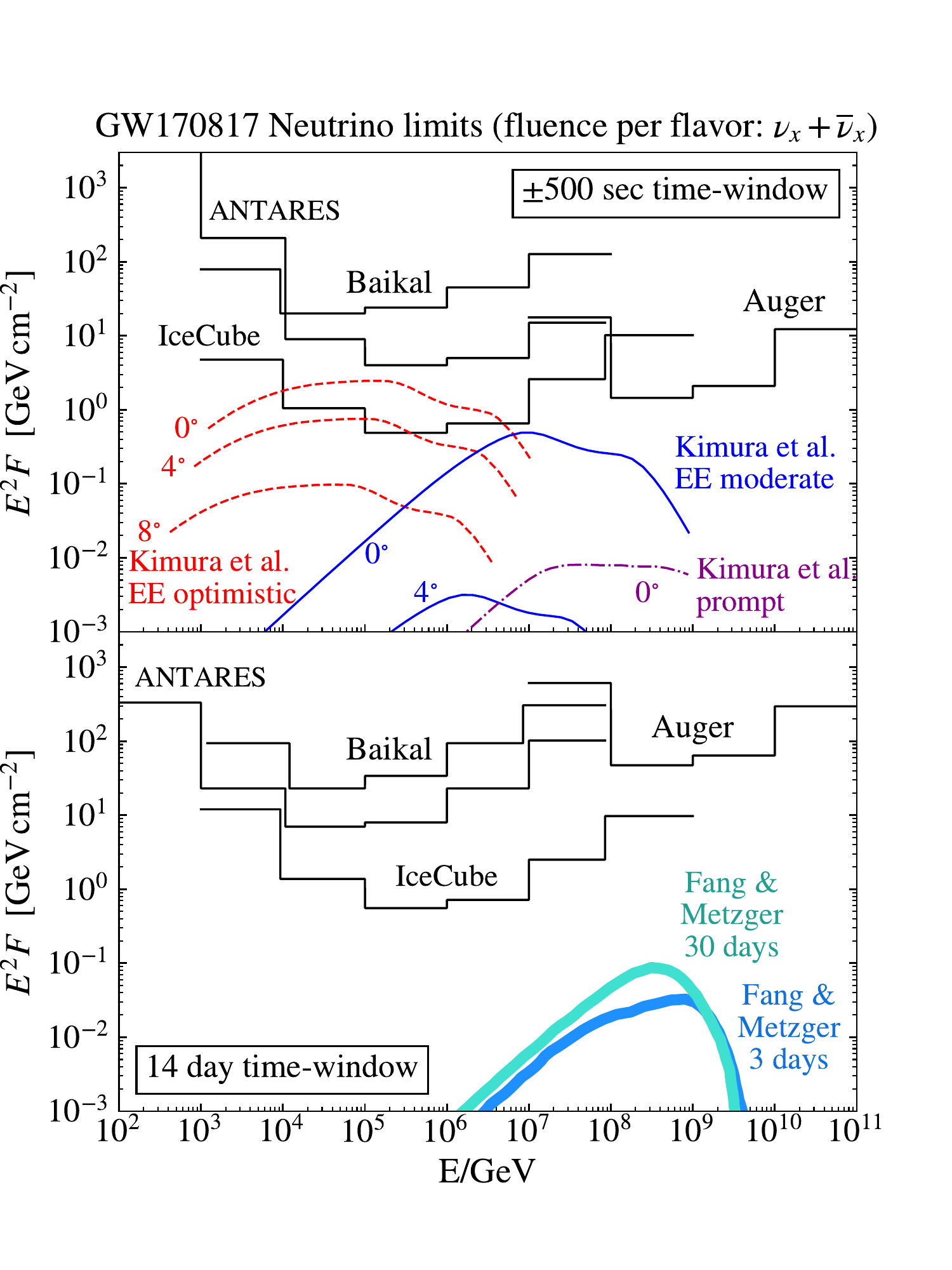}
\caption{High-energy neutrino fluence upper limits as a function of neutrino energy for the binary neutron star merger GW170817, based on data from ANTARES, IceCube, and the Pierre Auger Observatory. (a) Limits for a $\pm$500 s time window around the merger. (b) Limits over a 2-week period. Several model predictions are shown for comparison~\cite{2017ApJ...848L...4K,2017ApJ...849..153F}.
EE stands for extended emission. Adapted from Refs.~\cite{2017ApJ...850L..35A,2018arXiv181010966B}.
}
\label{fig:GW170817neutrino}
\end{figure}

Advanced LIGO/Virgo's second observing run was crowned by the multi-messenger discovery of the binary neutron star merger GW170817 a few days before the end of the run. This detection also represented a unique opportunity for high-energy neutrino searches. While no coincident neutrino was detected, the joint analysis of the participating observatories, ANTARES, IceCube and  Pierre Auger, were used to compute a joint constraint on neutrino emission from the merger over 9 orders of magnitude of energy, from 100\,GeV to 100\,EeV. These observational constraints, in comparison to selected emission scenarios, are shown in Fig.~\ref{fig:GW170817neutrino}. 
These results show that, assuming some of the optimistic emission models from short GRBs~\cite{2017ApJ...848L...4K}, we can rule out on-axis emission of optimistic scenarios related to extended emission, which is consistent with the large viewing angle inferred both from the gravitational-wave data and from afterglow observations~\cite{2018arXiv180511579T,2018Natur.561..355M}. 
Although the current afterglow observations are not enough to determine the entire jet structure~\cite{2019MNRAS.487.4884I}, the detection of GW170817 also indicated that high-energy emission may be observable at greater viewing angles than previously anticipated~\cite{2018arXiv180806238G,Murase:2017snw,2018PhRvD..98d3020K}, making nearby binary mergers an interesting target for coming joint observing periods.

\section{Source Models}
\label{sec:models}

\begin{figure}[th]
\includegraphics[width=3in]{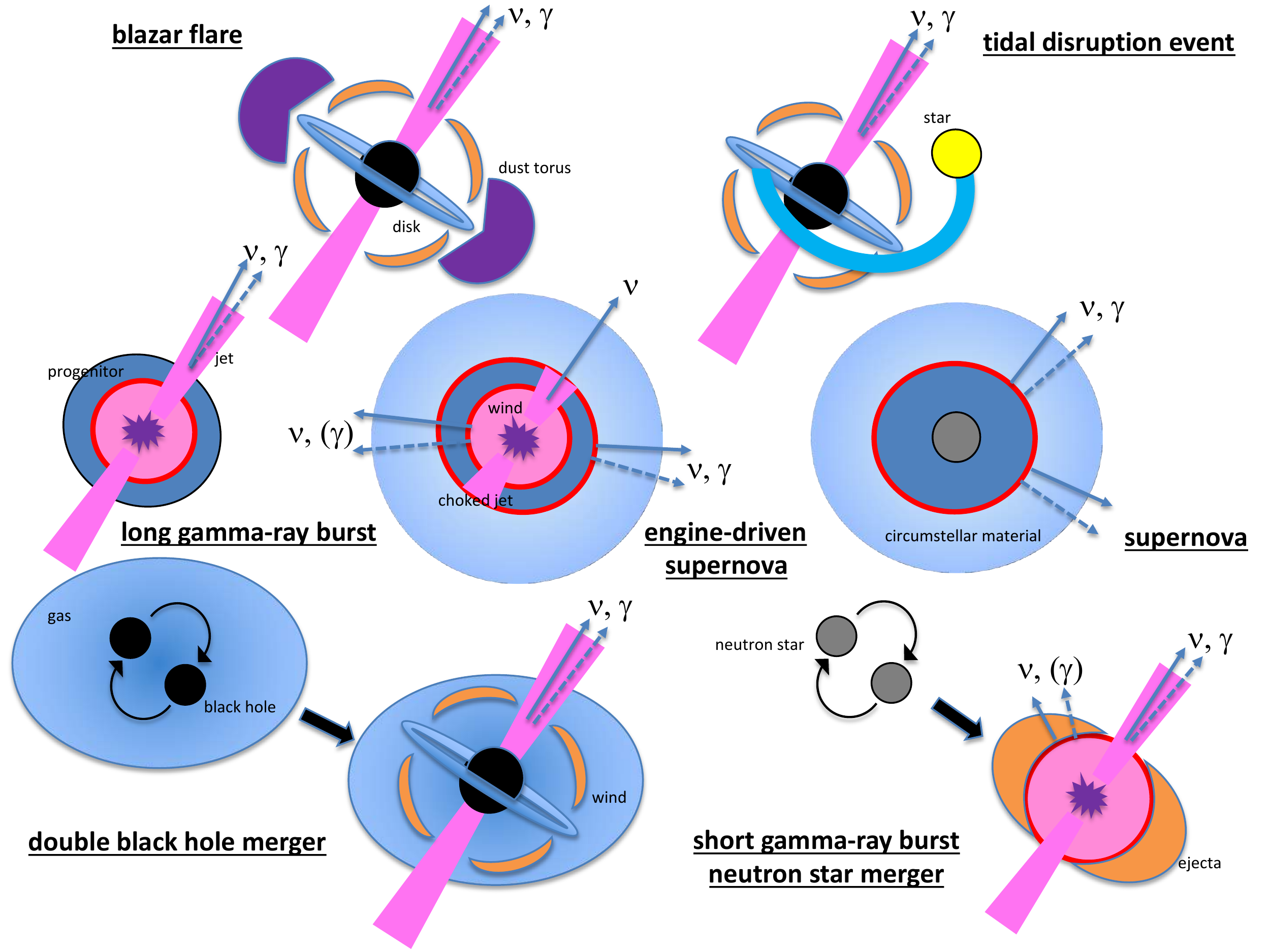}
\caption{Schematic picture of various high-energy multi-messenger transients.}
\label{figtransient}
\end{figure}

In this section, we discuss several possible sources of neutrinos and gravitational wave sources, which can be accompanied by high-energy emissions. High-energy emission mechanisms are schematically shown in Figure~\ref{figtransient}, and the list of these sources is summarized in Table~1 with some characteristic numbers. 

\begin{table}[th]
\tabcolsep7.5pt
\caption{List of multi-messenger transients that can be promising emitters of high-energy neutrinos and/or gravitational waves.}
\label{tab1}
\begin{center}
\begin{tabular}{@{}l|c|c|c|c@{}}
\hline
Source & Rate density & EM Luminosity & Duration & Typical Counterpart\\
 & $[{\rm Gpc}^{-3}~{\rm yr}^{-1}]$ & $[{\rm erg}~{\rm s}^{-1}]$ & $[\rm s]$ & \\
\hline
Blazar flare$^{\rm a}$ & $10-100$ & ${10}^{46}-{10}^{48}$ & ${10}^6-{10}^7$ & broadband \\
\hline
Tidal disruption event & $0.01-0.1$ & ${10}^{47}-{10}^{48}$ & $10^6-10^7$ & jetted (X)\\
& $100-1000$ & ${10}^{43.5}-{10}^{44.5}$ & $>10^6-10^7$ & tidal disruption event (optical,UV)\\
\hline
Long GRB & $0.1-1$ & ${10}^{51}-{10}^{52}$ & $10-100$ & prompt (X, gamma)\\
Short GRB & $10-100$ & ${10}^{51}-{10}^{52}$ & $0.1-1$ & prompt (X, gamma)\\
Low-luminosity GRB & $100-1000$ & ${10}^{46}-{10}^{47}$ & $1000-10000$ & prompt (X, gamma)\\
GRB afterglow & & $<{10}^{46}-{10}^{51}$,$$ & $>1-10000$ & afterglow (broadband)\\
\hline
Supernova (II) & ${10}^5$ & ${10}^{41}-{10}^{42}$ &$>{10}^5$ & supernova (optical) \\
Supernova (Ibc) & $3\times{10}^4$ & ${10}^{41}-{10}^{42}$ &$>{10}^5$ & supernova (optical) \\
Hypernova & $3000$ & ${10}^{42}-{10}^{43}$ &$>{10}^6$ & supernova (optical) \\
\hline
NS merger  & $300-3000$ & ${10}^{41}-{10}^{42}$ & $>10^5$ & kilonova (optical/IR) \\
 &  & ${10}^{43}$ & $>{10}^7-{10}^8$ & radio flare (broadband) \\
\hline
BH merger & $10-100$ & $?$ & $?$ & $?$\\
\hline
WD merger & ${10}^4-{10}^5$ & ${10}^{41}-{10}^{42}$ & $>10^5$ & merger nova (optical)\\
\hline
\end{tabular}
\end{center}
\begin{tabnote}
$^{\rm a}$Blazar flares such as the 2017 flare of TXS 0506+056 are assumed for the demonstration.\\ 
Abbreviations: BH, black hole; EM, electromagnetic; GRB, gamma-ray burst; NS, neutron star; WD, white dwarf.
\end{tabnote}
\end{table}

\subsection{Blazar Flares}
In general, blazars are highly variable objects that show broadband spectra from radio, optical, X-ray, and gamma-rays. 
In the standard leptonic scenario for SEDs, the low-energy and high-energy humps are explained by synchrotron emission and inverse-Compton radiation from non-thermal electrons, respectively. 
For BL Lac objects that typically belong to a low-luminous class of blazars, seed photons for the inverse-Compton scattering are mainly supplied by the electron synchrotron process. 
In contrast, flat-spectrum radio quasars (FSRQs) tend to be more luminous; it is believed that the external inverse Compton process is important for FSRQs. The origin of external target photon fields is under debate, which may come from an accretion disk, broad-line regions, a surrounding dusty torus, and the sheath region of a structured jet.

The high-energy hump could be dominated by a hadronic component, which is the so-called lepto-hadronic scenario. The gamma rays can be attributed to either cosmic-ray--induced electromagnetic cascade emission or ion synchrotron radiation~\cite{2019Galax...7...20B}. In the former case, the lepto-hadronic scenario predicts that the gamma-ray flux is comparable to the neutrino flux. The latter case usually requires strong magnetic fields, and does not necessarily accompany efficient neutrino production.      

In either scenario for the explanation of high-energy gamma rays, it is reasonable to consider a hybrid picture, in which both cosmic-ray ions and electrons are co-accelerated in the source. Target photons are not only synchrotron photons from primary leptons but also external radiation fields~~\cite{Murase:2014foa,Dermer:2014vaa}. As an example, let us consider a scattered accretion disk field. The effective optical depth to $p\gamma$ interactions is given by
\begin{equation}
f_{p\gamma}\approx \hat{n}_{\rm ext}\sigma_{p\gamma}^{\rm eff}r_{\rm ext}\sim0.01~\left(\frac{\tau_{\rm sc}}{0.1}\right){\left(\frac{r_{\rm sc}}{10^{18}~{\rm cm}}\right)}^{-1}{\left(\frac{L_{\rm AD}}{10^{46.5}~{\rm erg}~{\rm s}^{-1}}\right)}{\left(\frac{\varepsilon_{\rm AD}}{10~{\rm eV}}\right)}^{-1}
.\label{fpgamma}
\end{equation}
where $\tau_{\rm sc}$ is the optical depth to Thomson scattering for disk photons, $r_{\rm sc}$ is the size of the scattering region, $L_{\rm AD}$ is the radiation luminosity of the accretion disk, and $\varepsilon_{\rm AD}$ is the typical energy of the disk photons.

\begin{figure}[th]
\includegraphics[width=3in]{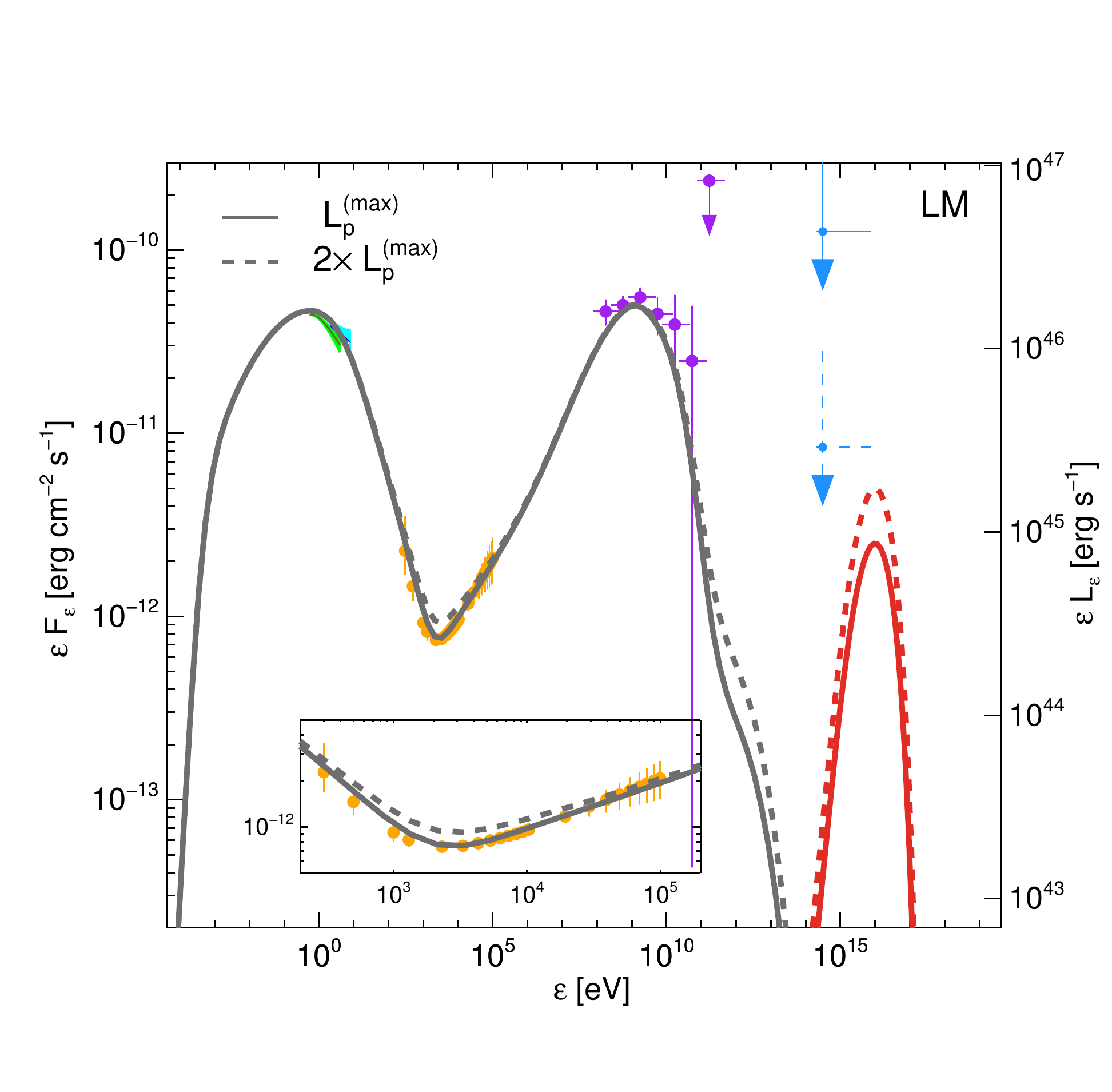}
\caption{Spectral energy distribution of TXS 0506+056 during the flare~\cite{Keivani:2018rnh}. 
The neutrino flux estimated by real-time alerts is from Ref.~\cite{Aartsen2018blazar1}. LM stands for Leptonic Model.}
\label{figTXS}
\end{figure}

The detection of IceCube-170922A, associated with TXS 0506+056, provided new insight into the origin of gamma rays. Here electromagnetic cascades inside the source play a crucial role in extracting implications for the source physics~\cite{Keivani:2018rnh,Ahnen:2018mvi,Murase:2018iyl,Cerruti:2018tmc,Gao:2018mnu}. 
High-energy neutrinos provide a smoking gun of cosmic-ray acceleration, so one naively expects that the neutrino detection would support the lepto-hadronic scenario. However, this is not the case. The SED of this blazar clearly showed the peak below $3\times{10}^{14}$~Hz and the dip in the X-ray range~\cite{Keivani:2018rnh}, which strongly constrains hadronic components. The neutrino flux is basically limited by the X-ray flux, as shown in Figure~\ref{figTXS}. Thus, proton-induced cascades cannot give a viable explanation for gamma rays. The proton synchrotron emission can explain the gamma-ray component, but the neutrino flux in the $0.1-1$~PeV range is predicted to be too low to explain the best-fit flux level of the IceCube data. Thus, ironically, the leptonic scenario is supported if IceCube-170922A originates from the flare of this blazar.

Besides, the fact that gamma rays were detected by MAGIC implies that the effective optical depth to $p\gamma$ interactions has to be very small, i.e., the required cosmic-ray power is too large. The cascade problem is even more serious for the past neutrino flare event in 2014-2015~\cite{Murase:2018iyl,Reimer:2018vvw,Rodrigues:2018tku}. These challenges may indicate the necessity of multi-zone models~\cite{Murase:2018iyl}. 

 \subsection{Tidal Disruption Events}
A star can be swallowed by a supermassive black hole located in the center of a galaxy. While the star is approaching the black hole, it can be tidally disrupted by the gravitational force, which occurs at the tidal disruption radius. 
About half of the mass is ejected, whereas the other forms an accretion disk and eventually falls back into the black hole. 
It is believed that the accretion initially proceeds as a super-Eddington mode and then becomes sub-Eddington. Resulting transients are observed as tidal disruption events (TDEs). 

Some of the TDEs posses relativistic jets that can be launched from the black-hole--accretion-disk system. Swift J1644+57 is thought to be such a jetted TDE. Strong non-thermal X rays were observed, with a typical duration of $t_{\rm dur}\sim10^6$~s. The bolometric radiation energy is ${\mathcal E}_\gamma\sim{10}^{54}$~erg, implying that the beaming corrected energy is ${\mathcal E}_j\sim10^{51}-10^{52}$~erg. Theoretically, it is widely discussed that the jets are powered by the Blandford-Znajek mechanism~\cite{Blandford:1977ds}.   

Cosmic-ray acceleration in TDEs was proposed by Ref.~\cite{Farrar:2008ex}, as a ``giant flare'' scenario, and associated neutrino emission has also been calculated~\cite{Murase:2008zzc,Wang:2011ip,Dai:2016gtz,Senno:2016bso,Lunardini:2016xwi}. 
The discovery of Swift J1644+57 revealed that jetted TDEs are strong X-ray sources~\cite{2011Natur.476..421B}. High-energy protons efficiently interact with these X rays. Equation~(\ref{eq:fpgamma}) infers that the effective $p\gamma$ optical depth is
\begin{equation}
f_{p\gamma}[\varepsilon_p]\sim1\frac{(L_\gamma^b/10^{47.5}~{\rm erg}~{\rm s}^{-1})}{(r/10^{14.5}~{\rm cm}){(\Gamma/10)}^2(\varepsilon_\gamma^b/1~\rm keV)}{\left(\frac{\varepsilon_{p}}{\varepsilon_{p}^b}\right)}^{\beta-1},
\end{equation}
where $L_\gamma^b$ is the luminosity at the peak energy $\varepsilon_\gamma^b$, and $\beta$ is the photon index. 
The above equation implies that jetted TDEs can be efficient neutrino emitters given that cosmic rays are accelerated in the jet. 

Non-detection of high-energy neutrinos from Swift J1644+57 implies that energy carried by cosmic rays is less than $\sim30{\mathcal E}_\gamma$. The contribution to the diffuse neutrino flux is expected to be $\lesssim10$\%~\cite{Dai:2016gtz,Senno:2016bso}, which is consistent with the limit from the absence of high-energy neutrino multiplets~\cite{Senno:2016bso}. 
If the disrupted star is a white dwarf, TDEs are expected to be promising gravitational wave sources~\cite{Kobayashi:2004py,Haas:2012bk}.

\subsection{Supernovae}
Massive stars with a stellar mass of $\gtrsim8M_\odot$ lead to the supernova explosion. During the gravitational collapse of a progenitor core, the central temperature increases and most of the gravitational binding energy is extracted by thermal neutrinos, which is estimated as ${\mathcal E}_{\rm G}\approx\left(\frac{GM_{\rm ns}^2}{R_{\rm ns}}\right)\sim 3\times{10}^{53}~{\rm erg}~{\left(\frac{M_{\rm ns}}{1~M_\odot}\right)}^2{\left(\frac{R_{\rm ns}}{{10}^{6}~{\rm cm}}\right)}^{-1}$, where $M_{\rm ns}$ is the remnant mass and $R_{\rm ns}$ is the radius.
Supernovae are known to be MeV neutrino emitters, as established by the SN 1987A detection~\cite{1992PhR...220..229K,1989NCimC..12...75A}.

High-energy neutrinos with energies beyond GeV or TeV can be produced in two ways. In the first case, cosmic rays are accelerated by a supernova shock, and the neutrinos are produced by their interactions with the ambient material. This situation is analogous to that of supernova remnants.  
The second possibility is that cosmic rays are supplied by outflows from the engine, which will be discussed further in subsections 4.4 and 4.5 with focuses on GRBs and engine-driven SNe.  

In the early stages of the supernova remnants, most the energy is in the kinetic form, and the energy fraction carried by cosmic rays is expected to be negligibly small. However, the recent observations of extragalactic supernovae have showed that significant mass losses ubiquitously occur before the explosion~\cite{Smith:2014txa}. The most spectacular examples are Type IIn supernovae, which have clear indications of interaction with the circumstellar material (CSM). Some of them, which are usually classified as Type IIn supernovae, indicated that the CSM mass reaches $M_{\rm cs}\sim1-10~M_\odot$ given that the CSM is spherical. Even Type II-P supernovae, which are most common among core-collapse supernovae, may have a significant CSM mass with $M_{\rm cs}\sim{10}^{-3}-10^{-1}~M_\odot$. 

\begin{figure}[h]
\includegraphics[width=3in]{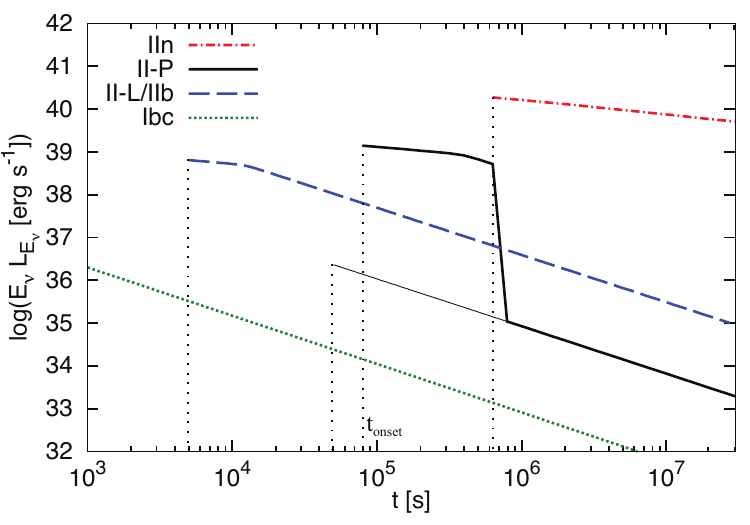}
\caption{High-energy neutrino light curves expected for various types of core-collapse supernovae (at 1~TeV). Adapted from Ref.~\cite{Murase:2017pfe}.}
\label{figSN}
\end{figure}

As the shock propagates, photons eventually break out, and then the shock becomes collisionless and is not mediated by radiation. Then one may expect the diffusive shock acceleration mechanism to operate as in supernova remnants. 
The accelerated protons should interact with gas via $pp$ interactions, and the effective optical depth for inelastic $pp$ interactions is estimated to be
\begin{equation}
f_{pp}\approx\kappa_{pp}\sigma_{pp}(\varrho_{\rm cs}/m_H)r_{s}(c/v_s)\sim1~(M_{\rm cs}/{10}^{-2}~M_\odot){(r_s/10^{14}~{\rm cm})}^{-2}{(v_s/3000~{\rm km}~{\rm s}^{-1})}^{-1},
\end{equation}
where $r_s$ is the shock radius, $v_s$ is the shock velocity, and $\varrho_{\rm cs}$ is the CSM density. This equation implies that high-energy neutrino and gamma-ray production efficiently occurs at early times. Neutrino light curves for various types of supernovae are shown in Figure~\ref{figSN}. IceCube can detect $\sim100-1000$ high-energy neutrinos from a Type II-P supernova~\cite{Murase:2017pfe}, if the next Galactic supernova occurs at $d\sim10$~kpc. Detection of high-energy emission from extragalactic supernovae requires stronger CSM interactions, which can be expected for Type IIn supernovae~\cite{Murase:2010cu,Petropoulou:2017ymv}. Searches for GeV-TeV gamma-ray emission have also been performed but the constraints are still consistent with theoretical predictions~\cite{2014ApJ...780...21M,TheFermiLAT:2015kla,2019ApJ...874...80M}. 

Some of Type Ibc supernovae with a relativistic velocity component -- transrelativistic supernova that are often associated with low-luminosity GRBs --- can also be neutrino and gamma-ray emitters owing to interactions with the CSM. See Refs.~\cite{Katz:2011zx,Kashiyama:2012zn} for more details. 

Core-collapse supernovae represent one of the promising directions of gravitational-wave studies~\cite{2001ApJ...550..372F,2008PhRvD..78f4056D,2012PhRvD..86b4026O,2018arXiv181207703R}. As core-collapse events are hidden from electromagnetic observations by the stellar material, only gravitational waves and thermal MeV neutrinos are able to carry information directly from the collapse to observers. Nevertheless, most emission models and numerical simulations predict gravitational-wave emission that is detectable by Advanced LIGO/Virgo for core-collapse supernovae within the Milky Way~\cite{2009CQGra..26f3001O,2006RPPh...69..971K}. Without a rapidly rotating core, gravitational-wave frequency will be characteristic to the newly formed protoneutron star's oscillation frequencies, while the gravitational-wave amplitude may be characteristic of the accretion rate~\cite{2018arXiv181207703R}. Much stronger gravitational-wave emission can be produced by rapidly rotating cores, in which dynamical and dissipative instabilities can result in a rotating non-axisymmetric structure that can radiate away some of the protoneutron star's angular momentum in gravitational waves~\cite{2009CQGra..26f3001O}. The amount of angular momentum available for gravitational wave radiation can be further increased by fallback accretion~\cite{2012ApJ...761...63P}. If the conversion of angular momentum is efficient, gravitational waves from core collapse supernovae with rapidly rotating cores can be detected out to tens of megaparsecs~\cite{Cutler:2002nw,Stella:2005yz,Kashiyama:2015eua}.


\subsection{Long Gamma-Ray Bursts}
\label{sec:LGRBs}
Long GRBs are among the brightest explosive astrophysical phenomena in the universe. Their isotropic-equivalent luminosities in gamma rays reach $L_{\rm iso}\sim{10}^{51}-{10}^{52}~{\rm erg}~{\rm s}^{-1}$ with a duration of $t_{\rm dur}\sim10-100$~s. These observations imply that the isotropic-equivalent gamma-ray energy is ${\mathcal E}_{\rm iso}\sim{10}^{53}~{\rm erg}$. This value is comparable to the isotropic-equivalent kinetic energy of GRB jets, which is inferred by multi-wavelength observations of the GRB afterglow emission.  
The outflows are thought to be collimated, and the true energy of the jet is ${\mathcal E}_j\sim10^{51}$~erg if the jet opening able is $\theta_j\sim0.1$. 
The central engine of the GRB jets and properties of the jet are unknown. It is believed that the jet is powered by a black hole with an accretion disk or a strongly magnetized neutron star (so-called magnetar). In the former case, the energy budget is limited by the rotation energy of a spinning black hole, ${\mathcal E}_{\rm BH-rot}=\left[1-\sqrt{\frac{1+\sqrt{1-a_*^2}}{2}}\right]M_{\rm BH}c^2\sim4\times{10}^{53}~{\rm erg}~\left(\frac{M_{\rm BH}}{3~M\odot}\right)$,
where $a_*=a/M_{\rm BH}$ and $a_*\sim0.7$ is assumed in the last estimate. 
In the latter case, the energy source is rotation energy of the remnant star, which is given by ${\mathcal E}_{\rm NS-rot}=\frac{1}{2}I{\left(\frac{2\pi}{P_i}\right)}^2
\sim2\times{10}^{52}~{\rm erg}~{\left(\frac{M_*}{1.4~M_\odot}\right)}{\left(\frac{R_{*}}{10~{\rm km}}\right)}^{2}{\left(\frac{P_i}{1~{\rm ms}}\right)}^{-2}$, 
where $I\approx0.35M_{*}R_{*}^2$ is inertia of momentum, $M*$ is the stellar mass, and $R_*$ is the stellar radius. 
Neutrino and gravitational wave signals can provide us with precious information about the central engine and jet composition.  

\begin{figure}[h]
\includegraphics[width=3in]{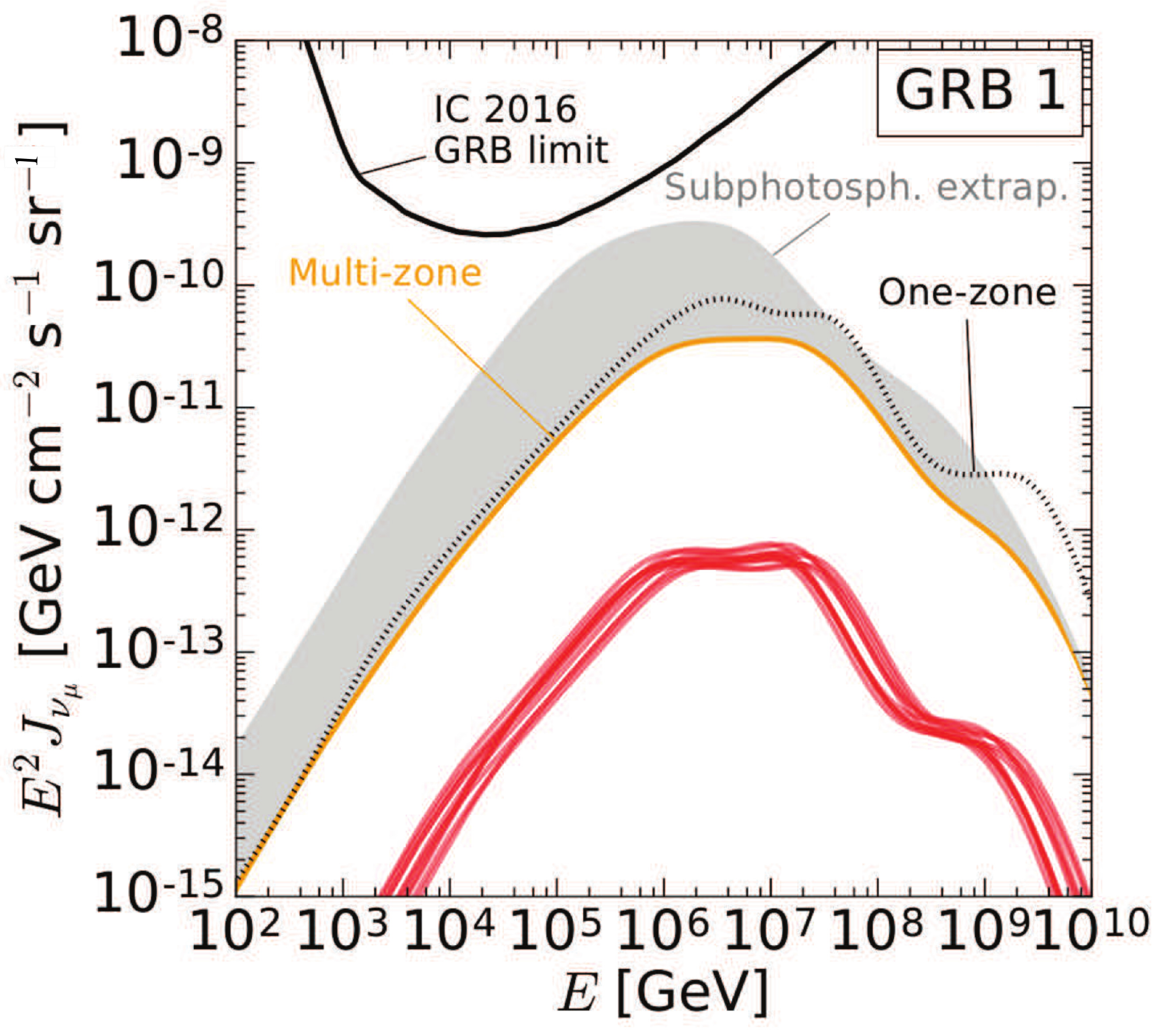}
\caption{Aggregated neutrino fluxes from gamma-ray burst (GRB) prompt emission. The figure shows the differential limit from IceCube as well as a possible contribution from subphotospheric neutrino emission. IC 2016 refers to the IceCube 2016 upper limit, which was calculated using their latest reported detector effective area and exposure in a stacked GRB search using tracks coming from the Northern Hemisphere~\cite{Aartsen:2016qcr}. Adapted from Ref.~\cite{Bustamante:2016wpu}.}
\label{figGRB}
\end{figure}

Prompt gamma-ray emission originates from internal dissipation in a relativistic jet with a bulk Lorentz factor of $\Gamma\sim100-1000$, and the gamma-ray energy spectrum has a peak around ${\varepsilon}_\gamma^b\sim1$~MeV.  
The emission mechanism has been under debate for many years. In the classical picture~\cite{2006RPPh...69.2259M,Kumar:2014upa}, the observed gamma rays are attributed to synchrotron radiation from non-thermal electrons that are accelerated inside a jet. Particles may be accelerated by internal shocks, which are thought to be mildly relativistic. However, efficient shock acceleration does not occur if the shock is relativistic and strongly magnetized, and magnetic reconnections are also considered as a promising mechanism. In either case, not only electrons but also ions will be accelerated in the jet, and high-energy neutrinos can be produced by $p\gamma$ interactions. Using equation~(\ref{eq:fpgamma}), the effective optical depth to $p\gamma$ interactions is estimated to be
\begin{equation}
f_{p\gamma}[\varepsilon_p]\sim0.01\frac{(L_\gamma^b/10^{51.5}~{\rm erg}~{\rm s}^{-1})}{(r/10^{14.5}~{\rm cm}){(\Gamma/10^{2.5})}^2(\varepsilon_\gamma^b/1~\rm MeV)}{\left(\frac{\varepsilon_{p}}{\varepsilon_{p}^b}\right)}^{\beta-1}.
\end{equation}
For GRBs, we have $\beta\sim1$ and $\beta\sim2-3$ for low- and high-energy spectral portions, respectively. The typical energy of neutrinos is predicted to be $0.1-1$~PeV~\cite{Waxman:1997ti}, which is the ideal energy range for IceCube. The importance of multi-pion production and other higher resonances has been investigated~\cite{Murase:2005hy,Baerwald:2010fk}. An example of the latest theoretical calculations~\cite{Bustamante:2016wpu} is shown in Figure~\ref{figGRB}. For GRB-like transients, stacking analyses are powerful, and the contribution to the diffuse neutrino flux is constrained to be less than $\sim1$\%~\cite{Aartsen:2014aqy,Bustamante:2014oka}. 
However, the possibility that GRBs are responsible for the observed UHECR flux has not been excluded yet, and further observations are necessary~\cite{Baerwald:2013pu,Globus:2014fka,Biehl:2017zlw}. 
Also, dimmer populations of bursts, such as low-luminosity GRBs, are missing in GRB samples used in the stacking analyses, so they can still make a significant contribution to the diffuse neutrino flux~\cite{Murase:2006mm,Gupta:2006jm,Senno:2015tsn,Tamborra:2015fzv} as well as the UHECR flux~\cite{Murase:2008mr,Zhang:2017moz,Boncioli:2018lrv,Zhang:2018agl}. 

Alternatively, the observed prompt gamma rays can be attributed to quasi-thermal, photospheric radiation. 
There are various photospheric models in the literature~\cite{2006RPPh...69.2259M,Kumar:2014upa}. Invoked subphotospheric dissipation mechanisms include internal shocks, magnetic reconnections, and collisions with neutron-loaded outflows. 
Whereas high-energy neutrino production around the photosphere is possible~\cite{Murase:2008sp,Wang:2008zm,Zhang:2012qy}, efficient acceleration of cosmic rays at shocks deep inside the photosphere is unlikely when the hydrodynamical shock is collisional or mediated by radiation~\cite{Murase:2013ffa}. (A strong subshock is in principle possible if the shock is magnetized, but the cosmic-ray ion acceleration is inefficient for strongly magnetized, perpendicular shocks.)
On the other hand, even if cosmic-ray acceleration does not occur, neutrinos can naturally be produced by neutrons. It is natural that neutrons are entrained into the jet. The neutrons are initially coupled with protons. But the decoupled neutron flow is eventually caught by a faster flow, causing inelastic $np$ collisions~\cite{Bahcall:2000sa,2010MNRAS.407.1033B}. Or internal shocks between compound flows are also accompanied by the dissipation via inelastic collisions. In either case, quasi-thermal neutrinos are expected, and the typical energy of neutrinos is $\varepsilon_\nu\sim10-100$~GeV. 
These neutrinos can be detected by more detailed analyses using the DeepCore data~\cite{Murase:2013hh,2013PhRvL.110x1101B}.  

Despite their typically large distance from the Earth, the progenitors of some long gamma-ray bursts may produce detectable gravitational waves. This requires that the collapsing stellar core that will produce the gamma-ray burst first forms a rapidly rotating protoneutron star. Some of the protoneutron stars formed in stellar core collapse may survive sufficiently long to develop dynamical or even dissipative instabilities, which deform the protoneutron star and result in the emission of gravitational waves. If a significant fraction of the protoneutron star's rotational energy can be converted to gravitational waves, this emission could be detectable out to tens of megaparsecs with Advanced LIGO/Virgo \cite{2013CQGra..30l3001B}. Very massive stars, however, can collapse without a supernova explosion|the so-called collapsar scenario|leaving virtually no time for a protoneutron star to form and emit gravitational waves. Possible gravitational-wave emission in this scenario may come from the fragmentation of the accretion disk \cite{Kobayashi:2002by,2007ApJ...658.1173P,2011PhRvL.106y1102K} or the collapsing star \cite{2002ApJ...565..430F,Ott:2010gv} or anisotropic neutrino emission~\cite{Suwa:2009si} or GRB jets~\cite{Sago:2004pn}, although these emission processes are currently uncertain. Alternatively, some long gamma-ray bursts may be produced directly by rapidly-rotating protoneutron stars with strong magnetic fields instead of a black hole--accretion disk system.
In this scenario the protoneutron star survives for a longer time, is fast rotating and is accreting additional matter, all favoring gravitational wave emission~\cite{2009ApJ...702.1171C}. 

\subsection{Engine-Driven Supernovae}
GRBs are caused by a relativistic jet that successfully breaks out from the progenitor star. However, the jet will not necessarily penetrate. It is natural for a sufficiently low-power jet to get ``choked'' inside a progenitor or dense CSM~\cite{Meszaros:2001ms,Razzaque:2004yv,Ando:2005xi,Iocco:2007td}. 
Such failed GRBs may be observed as ``engine-powered'' supernovae. Some of them are thought to become low-luminosity GRBs, whose properties are intermediate between supernovae and GRBs. Indeed such objects have been observed, in which the jet marginally fails or succeeds and a trans-relativistic component is seen in the ejecta velocity distribution.  
They are likely to be more ubiquitous than canonical high-luminosity GRBs. The gamma-ray emission mechanism is under debate, and the most popular scenario is that it originates from shock breakout of the relativistic ejecta in a dense CSM~\cite{Campana:2006qe}. As discussed above, high-energy neutrino and gamma-ray emission may occur around the shock breakout.  

Choked jets embedded in the stellar material or CSM are promising sources of high-energy neutrinos, given that cosmic rays are accelerated by the jets. Importantly, the system is calorimetric in the sense that sufficiently high-energy cosmic rays are depleted for neutrino and gamma-ray production, i.e., ${\rm min}[1,f_{p\gamma}]\approx1$.
The emitted neutrinos are called orphan neutrinos (if the jet is deeply choked and little gamma-ray emission is produced) and precursor neutrinos (if delayed gamma-ray emission is accompanied). 
However, cosmic-ray acceleration is suppressed when the shock is radiation mediated. Radiation largely smears the upstream structure, leading to a much weaker subshock, and energy carried by low-energy cosmic rays becomes small. This ``radiation constraint'' suggests that canonical high-luminosity GRBs are unlikely to be emitters of high-energy neutrinos~\cite{Murase:2013ffa}. 
Low-power GRBs, which can be produced if the jet is intrinsically weak and/or if the stellar material is extended, allow the cosmic-ray acceleration and associated neutrino production. They are also suggested as the main sources of high-energy neutrinos in the 10-100 TeV range~\cite{Murase:2013ffa,Senno:2015tsn,Denton:2017jwk}. As noted above, these medium-energy neutrinos suggest the existence of hidden neutrino sources.  

Energy injection from the central engine does not have to be caused by relativistic jets that are collimated outflows. 
Winds from a pulsar or accretion disk around a black hole can also power the ejecta and resulting observed emissions. 
In particular, a fast-rotating pulsar or magnetar has been actively discussed as the central engine for various types of supernovae as well as GRBs~\citep[e.g.,][]{Thompson:2004wi,2016ApJ...818...94K,Margalit:2018bje}. The long-lived pulsars or magnetars are also intriguing sources of gravitational waves, and high-energy counterpart searches have been of much interest. 

Pulsar winds are expected to be Poynting dominated, and can form a pair of forward and reverse shocks via interaction with the supernova ejecta. Pulsar wind nebulae such as the Crab nebula have broadband, non-thermal spectra from radio, optical, X-ray and gamma-rays. Detailed modeling of the non-thermal nebular emissions indicates that the plasma is carried by electron-positron pairs, and a significant fraction of the electron-positron pairs are accelerated around the termination shock. 

It is natural to expect embryonic pulsar wind nebulae are also efficient accelerators of electrons and positrons. Then, bright X-ray counterparts can be expected as month-to-year transients~\cite{Metzger:2013kia}. In particular, hard X rays serve as powerful probes of pulsar-driven supernovae~\cite{Murase:2014bfa,2016ApJ...818...94K}, but there has been no indication for candidate supernovae including super-luminous ones~\cite{Margutti:2017lyd}.   
Gamma rays have a larger penetration power, and strong gamma-ray emission in the GeV-TeV range is produced by upscatterings of supernova photons. GeV gamma rays are detectable up to nearby supernovae within 100~Mpc, which are potential targets for Fermi-LAT~\cite{Murase:2014bfa} and searches have been performed~\cite{2018A&A...611A..45R}. 
TeV gamma-ray counterparts are interesting targets for imaging atmospheric Cherenkov telescopes such as MAGIC, VERITAS, HESS, and CTA. But they are subject to intrasource attenuation by supernova photons. 

Some ions can potentially be accelerated in the wind or around the termination shock. Even acceleration to ultrahigh energies has been suggested~\cite{Blasi:2000xm,Arons:2002yj}. Although details of ion acceleration by embryonic pulsar wind nebulae are unknown, possible mechanisms include surfing or wake-field acceleration. The ultrahigh-energy ions escaping from the nebula are damped in the ejecta and radiation field, and high-energy neutrinos are produced via both $pp$ and $p\gamma$ interactions~\cite{Murase:2009pg,Fang:2013vla}. For example, the effective optical depth to $pp$ interactions is estimated to be
\begin{equation}
f_{pp}\approx\kappa_{pp}\sigma_{pp}(\varrho_{\rm ej}/m_H)r_{\rm ej}
\simeq4~(M_{\rm ej}/M_\odot){(r_{\rm ej}/10^{15}~{\rm cm})}^{-2},
\end{equation}
where $r_{\rm ej}$ is the shock radius, $v_{\rm ej}$ is the shock velocity, and $\varrho_{\rm ej}$ is the ejecta density.
The system is calorimetric at early times. However, because of a high density of the ejecta, pions and muons are cooled before they decay, so that the neutrino flux can initially be suppressed at the highest energies. At late times, although the suppression is negligible, the neutrino flux declines following the spin-down power. 
Recently, the model has been applied to the fast blue optical transient, AT2018cow~\cite{Fang:2018hjp}.  

\subsection{Short Gamma-Ray Bursts and Neutron Star Mergers}
The connection between short gamma-ray bursts and neutron star mergers has been anticipated for decades~\cite{2006RPPh...69.2259M,1986ApJ...308L..43P,1989Natur.340..126E,2014ARA&A..52...43B}, and was strongly supported by the multi-messenger discovery of the neutron star merger GW170817 and its gamma-ray burst counterpart GRB\,170817A, although the origin of the prompt gamma rays is still under debate~\cite{2017ApJ...848L..13A}. As the two neutron stars approach each other during the merger, some of their mass gets tidally disrupted, forming a disk around the newly formed, central compact object. The central object would eventually collapse into a black hole. Accretion onto the black hole from the surrounding disk then drives a relativistic outflow.

The merger of a neutron star and a black hole can also produce similar relativistic outflows to those of neutron star mergers, but only if the black hole’s mass is sufficiently small ($\lesssim10$\,M$_\odot$) to tidally disrupt the neutron star before merging~\cite{2013CQGra..30l3001B,1999ApJ...527L..39J,2005ApJ...634.1202R,2012PhRvD..86l4007F,Kyutoku:2013wxa,Kyutoku:2015gda,2015PhRvD..92f4034K,2015ApJ...806L..14P,2018PhRvD..98l3017R}. No neutron star-black hole merger has been detected so far with gravitational waves, constraining their rate to $\lesssim 600$\,Gpc$^{-3}$yr$^{-1}$~\cite{2018arXiv181112907T}. The properties of relativistic outflows from neutron star-black hole mergers may be different from those of binary neutron star mergers due to the different black hole and ejecta masses of the two event types. In addition, a supramassive neutron star that forms in neutron star mergers can alter the outflow if it survives longer than a few milliseconds. 

Neutron star mergers are among the most promising sources of gravitational waves for Earth-based interferometers such as LIGO/Virgo. At Advanced LIGO/Virgo's design sensitivity, they will be detectable out to about 200\,Mpc on average~\cite{2016LRR....19....1A}, corresponding to a detection rate of $3-100$ per year~\cite{2018arXiv181112907T}. Gravitational waves will confirm which nearby high-energy event resulted from neutron star mergers. In addition, Advanced LIGO/Virgo will provide the masses of the merging black holes, which in turn can be used to determine how much neutron star matter got tidally disrupted and how long the newly formed supramassive neutron star is expected to live before collapsing into a black hole. Even more can be learned by jointly using information from gravitational waves and the detected electromagnetic/neutrino emission~\cite{Shibata:2017xdx,2017ApJ...850L..19M,2018arXiv181204803C}. Gravitational waves will also help constrain the equation of state of supranuclear matter~\cite{2018PhRvL.121p1101A}. Finally, gravitational waves carry information on the luminosity distance of neutron star mergers, which, together with the redshift of the merger's host galaxy provides an alternative distance ladder to constrain Hubble's constant~\cite{2017Natur.551...85A}.

Short gamma-ray bursts produced by neutron star mergers can be distinguished from long gamma-ray bursts produced by stellar core collapse by their duration, which is typically less than 2\,s, and their spectral hardness compared to the softer long gamma-ray bursts. While the two gamma-ray burst types have comparable peak luminosities, due to their durations, short gamma-ray bursts emit much less overall isotropic-equivalent energy in gamma rays, mostly within $10^{49}-10^{52}$\,erg~\cite{2014ARA&A..52...43B}. 

The discovery of binary neutron star merger GW170817 and gamma-ray burst counterpart GRB\,170817A strongly supported the existing short-GRB paradigm, while also providing interesting new questions for our understanding of the short GRB engine. In particular, the inclination of the orbiting binary was $15^\circ-40^\circ$ off the direction of Earth~\cite{2018arXiv180511579T}. The observation of gamma rays at such high inclination angle meant that the relativistic outflow is structured, with a stronger, highly beamed component along the inclination axis, and a weaker emission that extends to higher angles~\cite{2018PhRvL.120x1103L,2018Natur.561..355M,2017ApJ...848L..20M,2018PTEP.2018d3E02I}. The origin of this observed structure is not yet clear, one possibility being the interaction of the relativistic outflow with the lower velocity, quasi-isotropic dynamical and wind ejecta~\cite{2018MNRAS.479..588G,2018Natur.554..207M}.

Short GRBs may be important sources of high-energy neutrinos, with neutrino fluxes possibly comparable to the flux of gamma-rays, reaching up to $\sim10^{51}$\,erg of isotropic-equivalent energy~\cite{2017ApJ...848L...4K,Biehl:2017qen}. Neutrino emission can be even higher if gamma rays are partially attenuated, e.g., by the dynamical ejecta surrounding the merger, which the relativistic outflow must burrow through~\cite{2018PhRvD..98d3020K}. 
As the beamed outflow from neutron star mergers is expected to be neutron rich, the collision of relativistic protons with slower neutrons also represents an alternative mechanism to convert the outflow's kinetic energy to gamma-rays and GeV neutrinos~\cite{Murase:2013hh,2013PhRvL.110x1101B}.

GeV-TeV gamma-ray emission from GRB 170817A has been searched for but no positive signal was found~\cite{Fermi-LAT:2017uvi,Abdalla:2017mtd}
It is known that some short GRBs are accompanied by extended and plateau emissions. These photons can be upscattered by relativistic electrons accelerated at the jet, and the resulting GeV-TeV gamma rays could be detected by gamma-ray telescopes~\cite{Murase:2017snw}. In particular, CTA is expected to be powerful for long-lasting gamma-ray counterpart searches.

\subsection{Black Hole Mergers}
Stellar-mass binary black hole mergers represent the primary source of gravitational waves, with detection rates that could reach one per day within the next years~\cite{2018arXiv181112907T,2016LRR....19....1A}. Binary black holes may originate from either binary stellar systems that both undergo stellar collapse~\cite{2002ApJ...572..407B}, or from dynamical encounters in galactic nuclei or globular clusters~\cite{2009MNRAS.395.2127O,2015PhRvL.115e1101R,2018Natur.556...70H,2018PhRvL.121p1103F,2006MNRAS.372.1043I}. These different formation channels result in different binary properties, such as mass, mass ratio and spin.

Binary black hole mergers are generally not expected to result in emission other than gravitational waves. However, some of the binaries may merge in dense environments in which sufficient gas is available for accretion to produce detectable electromagnetic or neutrino emission. The observation of a possible short GRB by the Gamma-ray Burst Monitor on the Fermi satellite in coincidence with the binary black hole merger GW150914 was a possible first hint for such an event \cite{2016ApJ...826L...6C} (but see Ref.~\cite{2016ApJ...827L..38G}). Scenarios that can result in electromagnetic and neutrino emission include mergers in the accretion disks of active galactic nuclei \cite{2017ApJ...835..165B,2017MNRAS.464..946S,2017NatCo...8..831B,2019arXiv190301405Y}, gas or debris remaining around the black holes from their prior evolution~\cite{2016ApJ...821L..18P,2016ApJ...822L...9M,2016ApJ...823L..29K,2016PhRvD..93l3011M} (but see Ref.~\cite{2017MNRAS.465.4406K}), and binary black hole formation inside a collapsing star~\cite{2016ApJ...819L..21L} (but see Ref.~\cite{2017MNRAS.470L..92D}). The electromagnetic and neutrino brightness of binary black hole mergers within these scenarios is currently not well constrained.

\subsection{White Dwarf Mergers}
Double white dwarf mergers are thought to be among the progenitors of Type Ia supernovae. However, details of violent merger processes are still under debate, and they may be observed as weaker optical transients~\cite{Beloborodov:2013kpa}. Numerical simulations suggested that the white dwarf mergers can result in the ejection of material with a mass of $\sim10^{-3}-{10}^{-2}~M_\odot$~\cite{Ji:2013sda}.    

The magnetic luminosity of the outflows is $L_B\sim10^{44}-10^{46}~{\rm erg}~{\rm s}^{-1}$, which can be accompanied by magnetic reconnections and particle acceleration. Following this scenario, one could expect high-energy neutrino emission from white dwarf mergers~\cite{Xiao:2016man}. Turbulence and efficient particle acceleration is expected beyond the photon diffusion radius. Considering the dissipation of magnetic energy via reconnections TeV-PeV neutrinos can be expected after the photons break out, and the signals may coincide with thermal emission in the optical band. 
These high-energy neutrinos can be used as probes of the outflow dynamics, magnetic energy dissipation, and cosmic-ray acceleration at subphotospheres. Note that the accompanying high-energy gamma rays are absorbed because of the large $\gamma\gamma$ optical depth, so these sources are among the hidden neutrino sources. 

Double white dwarf mergers are important targets for low-frequency gravitational wave observations with e.g., LISA~\cite{AmaroSeoane:2012je,Evans:1987qa,Korol:2017qcx}. Multi-messenger detections will enable us to probe the merger rate, binary formation and evolution mechanisms, and links to explosion mechanisms such as Type Ia supernovae.  

\section{Outlook}
\label{sec:outlook}

We presented a review of high-energy emission processes in cosmic transients in the context of multi-messenger observations. 
The era of these observations has just started, and we anticipate a rapidly growing number of such discoveries in the near future. This means that the field is set to develop and change in the near future. However, we believe that the present review can help guide the reader through well-established processes and where the interesting open questions currently lie. We summarize some of the main open questions below. 

The physical association between neutrinos and blazar flares is currently tentative, which should be confirmed by more discoveries with multi-messenger observations in the near future. Observational constraints from other blazars suggests that X-ray data are critical for the SED modeling and observational monitoring of blazar flares at multi-wavelengths, especially in the X-ray band. Stacking searches with IceCube data, based on more blazar flare samples, will also provide a complementary test.   
Theoretical predictions indicate that FSRQs are stronger emitters of high-energy neutrinos than BL Lac objects. The high-energy hump of the brightest FSRQs is expected to lie in the MeV range, and they are more common at higher redshifts. Thus, MeV observations with more sensitive telescopes such as AMEGO~\cite{Moiseev:2017mxg} will also be important, and the possibility that such blazars significantly contribute to the IceCube neutrino flux can be tested in future. Searches for ultrahigh-energy neutrinos in the EeV range are also important to test whether blazars are accelerators of UHECRs or not. 

Long GRBs and jetted TDEs are among the brightest X-ray and gamma-ray transients in the Universe. Even though they are not dominant in the diffuse neutrino sky, they are still viable as the main sources of UHECRs. Thus, further dedicated searches for neutrinos from GRBs and TDEs are necessary. Long GRBs are potential sources of gravitational waves, and TDEs are also expected to be intriguing gravitational wave emitters for the disruption of a white dwarf by an intermediate black hole. Coincidence searches with gravitational waves with next-generation gravitational-wave detectors such as the Einstein Telescope and Cosmic Explorer~\cite{Sathyaprakash:2012jk,Evans:2016mbw} will be crucial. 
We should also remark that relativistic jets of GRBs and TDEs propagate in the interstellar material and UHECR acceleration may occur in the afterglow phase. For such neutrino afterglows, the typical energy of neutrinos is expected in the EeV range, and observations with next-generation neutrino detectors such as the Askaryan Radio Array and Antarctic Ross Ice-Shelf ANtenna Neutrino Array that may merge into RNO (Radio Neutrino Observatory), GRAND (Giant Radio Array for Neutrino Detection), Trinity, and POEMMA (Probe Of Extreme Multi-Messenger Astrophysics), will be important~\cite{AlvesBatista:2019tlv}.   

Recent surveys in the optical and infrared bands revealed the diversity of supernovae, and some of the classes, such as super-luminous supernovae and hypernovae, may share a similar type of the central engine with GRBs and even fast radio bursts. Understanding the connections among these cosmic explosions is important to reveal the mechanisms of jets and outflows, and the roles of black holes and neutron stars. 
These types of explosions are promising sources of gravitational waves, and high-energy neutrinos and gamma rays will provide information about dense environments that cannot be probed by visible light. 
They might significantly contribute to the diffuse neutrino flux especially in the $10-100$~TeV range. 
Not only stacking analyses but also neutrino-triggered follow-up observations are encouraged to test the models. Neutrino observations with a sufficiently good angular resolution of $\sim0.1-0.2$~deg is necessary~\cite{Murase:2016gly}, which could be achieved by KM3Net~\cite{Adrian-Martinez:2016fdl} and IceCube-Gen2~\cite{Aartsen:2014njl}. 
Nearby supernovae, including the next Galactic supernova, are also interesting targets as multi-messenger sources. They are promising sources of MeV neutrinos and gravitational waves. In addition, high-energy neutrinos from Type II supernovae are detectable, and more than 100 events may be detected for the next Galactic event~\cite{Murase:2017pfe}. In this sense, supernovae can be not only multi-messenger but also multi-energy sources, and cosmic-ray ion acceleration may be observed in real-time by neutrino and gamma-ray observations. Not only MeV neutrinos but also GeV neutrinos might be seen by Hyper-Kamiokande~\cite{Hyper-Kamiokande:2016dsw}.     

In the next decade, we will have many events of gravitational wave signals from black hole and/or neutron star mergers. It is known that some short GRBs have extended and plateau emissions, so X-ray observations are important to understand the activities of the central engine. Regarding TeV gamma-ray searches, gamma-ray monitors such as HAWC (High Altitude Water Cherenkov Observatory) and SGSO (Southern Gamma-Ray Survey Observatory)~\cite{2019arXiv190208429A} will enable the observation of bright gamma-ray emissions during the prompt and early afterglow phases, whereas CTA (Cherenkov Telescope Array)~\cite{2019scta.book.....C} will play a role in deeper follow-up observations of gravitational wave transients. The coincident detection between high-energy neutrinos and gravitational waves from neutron star mergers may be challenging for the current IceCube but would be promising with next-generation neutrino detectors such as IceCube-Gen2.

\section*{DISCLOSURE STATEMENT}
The authors are not aware of any affiliations, memberships, funding, or financial holdings that might be perceived as affecting the objectivity of this review. 

\section*{ACKNOWLEDGMENTS}
The authors thank Markus Ahlers, Christopher Berry, Kunihito Ioka, Szabolcs Marka, Peter M\'esz\'aros, Christian Spiering, and the IceCube Collaboration. The article has been approved for publication by the LIGO Scientific Collaboration under document number LIGO−P1900117. The authors thank Pennsylvania State University and the University of Florida for their generous support. The work of K.M. is supported by the Pennsylvania State University, Alfred P. Sloan Foundation, and NSF grant No. PHY-1620777. IB is grateful for the generous support of the University of Florida and the National Science Foundation under grant PHY-1911796.





%

\section*{REFERENCES CITED}


\bibliographystyle{naturemag}
\bibliography{kmurase} 

\end{document}